\documentclass[lettersize,journal]{IEEEtran}
\usepackage[colorlinks,urlcolor=black,linkcolor=blue,citecolor=blue]{hyperref}

\def\BibTeX{{\rm B\kern-.05em{\sc i\kern-.025em b}\kern-.08em
    T\kern-.1667em\lower.7ex\hbox{E}\kern-.125emX}}

\ifCLASSOPTIONcompsoc
\usepackage[nocompress]{cite}
\else
  \usepackage{cite}
\fi

\usepackage{algorithm}
\usepackage{pgffor}
\usepackage{algpseudocode}
\usepackage{epsfig}
\usepackage{graphicx}
\usepackage{amsfonts}
\usepackage{comment}
\usepackage{color}
\usepackage{amssymb}
\usepackage{amsmath}
\usepackage{amsthm}

\usepackage{etoolbox}
\usepackage{color}
\usepackage{flushend}
\usepackage{babel,blindtext}
\usepackage{caption}
\usepackage{subcaption}

\usepackage{csquotes}
\usepackage{rotating}
\usepackage{amssymb}
\usepackage{algorithm}
\usepackage{algpseudocode}
\usepackage{color}
\usepackage{epsf}
\usepackage{psfrag}
\usepackage{epsfig}
\usepackage{graphicx}
\usepackage{amsfonts}
\usepackage{textcomp}
\usepackage{comment}
\usepackage{footnote}
\usepackage{tablefootnote}
\usepackage[flushleft]{threeparttable}
\usepackage{paralist}
\usepackage{mdwlist}
\usepackage{amssymb}
\usepackage{url}
\usepackage{verbatim}
\usepackage{alltt}
\usepackage{balance}
\usepackage{multirow}
\usepackage{epstopdf}
\usepackage{lipsum}
\usepackage{float}
\usepackage{booktabs}
\usepackage{slashbox}
\usepackage[table]{xcolor}

\usepackage{array}
\usepackage{boldline}
\usepackage{caption,setspace}
\usepackage{mathtools}
\usepackage{csquotes}

\usepackage{stackengine}[2013-10-15]

\usepackage[T1]{fontenc}
\usepackage{frcursive}
\usepackage{calligra}
\usepackage{wedn}
\usepackage{bm}
\usepackage{aurical}
\usepackage{upgreek}
\usepackage{pgffor}

\definecolor{green}{RGB}{0,204,0}

\usepackage{cite}
\usepackage{graphicx}
\usepackage{adjustbox}
\usepackage{graphicx}
\usepackage{subcaption}
\usepackage{esint}
\usepackage{tipa}

\usepackage{pgfplots}
\usepackage{pgfplotstable}
\pgfplotsset{compat=1.18}
\usetikzlibrary{patterns}

\newcolumntype{?}{!{\vrule width 1pt}}
\newcolumntype{+}{!{\vrule width 1.25pt}}

\def\hlineb#1{%
\noalign{\ifnum0=`}\fi\hrule \@height #1 %
\futurelet\reserved@a\@xhline}

\usepackage{cancel}

\definecolor{darkgreen}{RGB}{0,153,0}
\definecolor{darkred}{RGB}{192,0,0}

\ifCLASSINFOpdf
\else
\fi
\begin{document}







\title{Assessing Engineering Student Perceptions of Introductory CS Courses in an Indian Context}

\author{Utsav Kumar Nareti,~\IEEEmembership{Student Member,~IEEE},  
Divyansh Gupta, 
Chandranath Adak,~\IEEEmembership{Senior Member,~IEEE}, 
Soumi Chattopadhyay,~\IEEEmembership{Senior Member,~IEEE}, 
Emma Riese, 
Tanujit Chakraborty, 
Mayank Agarwal,
Satendra Kumar
\thanks{
U.K. Nareti, D. Gupta, C. Adak, M. Agarwal, and S. Kumar are with the Indian Institute of Technology Patna, Bihar 801106, India.
\\
S. Chattopadhyay is with the Dept. of CSE, Indian Institute of Technology Indore, Madhya Pradesh 453552, India. 
\\
E. Riese is with the School of EECS, KTH Royal Institute of Technology, SE-100 44 Stockholm, Sweden. 
\\
T. Chakraborty is with SAFIR, Sorbonne University Abu Dhabi, and SCAI, Sorbonne University, Paris, 75006, France.
\\
\emph{Corresponding author: C. Adak (email: chandranath@iitp.ac.in)}
}
}

\markboth{U. K. N\MakeLowercase{areti \textit{et al.}}}{}


\maketitle

\begin{abstract}
Understanding student perceptions of assessment is vital for designing inclusive and effective learning environments, especially in technical education. This study explores engineering students’ perceptions of assessment practices in an introductory computer science/ programming course, and its associated laboratory within an Indian engineering institute context. A total of 318 first-year Bachelor of Technology students participated in a weekly 25-statement Likert-scale survey conducted over nine weeks. Using descriptive statistics and non-parametric tests (Mann-Whitney U and Kruskal-Wallis), the analysis reveals that students largely perceive lab assignments as effective learning activities and view exams and projects as authentic and skill-enhancing. Students appreciated the role of instructors in shaping course content and found teaching assistants to be approachable and helpful, despite some inconsistencies. The study also finds significant variations in students’ academic performance and assessment perceptions based on prior programming experience, technology familiarity, gender, and academic branch. 
Notably, the performance data did not follow a Gaussian distribution, challenging common assumptions in grade modeling. A comparative analysis with European cohorts highlights both universal patterns and contextual differences, offering valuable insights for designing inclusive and equitable assessment strategies in programming education.

\end{abstract}


\section{Introduction}\label{sec:intro}

\IEEEPARstart{I}{ntroductory} programming, offered as the first Computer Science (CS) course along with its associated laboratory course, are mandatory requirements for many first-year {engineering} students, including both CS and non-CS majors. Introductory programming courses have been the focus of numerous studies, largely due to the common challenges students face in acquiring programming skills \cite{10.1145/3293881.3295779}. Difficulties in mastering the course are attributed not only to the inherent complexity of programming \cite{article1}, but also to social factors like a competitive classroom environment \cite{article}. Historically, the performance of students on course exams is often assumed to follow a normal distribution, reflecting that most are clustered around the mean and a small number either overachieve or underachieve \cite{gaultney2001grade}. Although the Gaussian assumption is the ubiquitous choice for modeling exam scores, there is empirical evidence of an asymmetric logit-normal assumption at the item level \cite{arthurs2019grades}. Computer programming is considered an essential requirement for all engineering students, providing key skills for academic and professional growth \cite{10036401}.

To manage the workload of large mandatory courses, course coordinators/ main instructors often work with teaching assistants (TAs), generally Master's or PhD students, to coordinate labs, lead tutorials, and assist with grading \cite{10036401}. While extensive research exists on CS courses, students' experiences with assessments remain underexplored \cite{10.1145/3293881.3295779}. This study addresses the gap by investigating how engineering students experience assessments in their introductory CS course and how they perceive the roles of course coordinators and TAs in these contexts. Furthermore, this study contributes with an Indian perspective to contrast the previous study results in a European setting \cite{10036401}. This group is particularly relevant, as many engineering programs require foundational programming skills, which are valuable even outside a CS major \cite{10.1145/3517134}.
Non-CS majors form a significant portion of the course enrollment and often do not view themselves as future programmers or computer scientists. Nevertheless, the skills acquired in this course are considered essential for broader engineering education and future career opportunities. 
Building on prior European research \cite{qualitative_step,10036401}, this study examines how Indian students' perceptions differ based on prior experience, gender, academic branch, and grades, and investigates how these patterns align with findings from international contexts. 
The central research questions (\emph{RQ}s) explored in this study are:


\emph{RQ1}. How do engineering students perceive and engage with different types of assessments, such as lab assignments, mid/end-term exams, and projects?

\emph{RQ2}. How do students view the roles and contributions of TAs and instructors during assessments?

\emph{RQ3}. Does the performance of students in the introductory CS course follow a Gaussian distribution?

\emph{RQ4}. What differences exist based on prior programming and technology experience, and how do factors such as gender, grades, and academic branch influence perceptions?

\emph{RQ5}. Are there any differences in perceptions or influencing factors between European and Indian educational contexts?


The rest of the paper is organized as follows: 
Section~\ref{sec:related_work} reviews background, Section~\ref{sec:method} details the methodology, Section~\ref{4sec:result} presents the results, and Section~\ref{5sec:discussion_conclusion} 
concludes the study with discussions and future research directions.

\section{Background}
\label{sec:related_work}


\subsection{Foundation and Context}
This study adopts a constructive learning approach \cite{cambridge}, wherein the course is developed from the ground up by integrating theoretical instruction with hands-on practice/ lab. Course coordinators emphasize foundational concepts to equip students for subsequent, related courses \cite{10.1145/3517134}. The assessment structure consists of two assignments, two quizzes, and two main examinations: midterm and endterm that hold greater weight in the overall evaluation. A course project assigned early in the term requires students to demonstrate running code with a brief presentation and Q\&A session \cite{10.1145/1565799.1565820}. The lab component is closely aligned with the main course, involving weekly structured three-hour sessions and assessments through weekly evaluations, as well as midterm and end-term lab exams comprising practical tasks and a viva. This format, common across many engineering institutions, fosters interest in programming among non-CS majors and underscores its relevance across engineering disciplines \cite{10.1145/961290.961519}. By addressing structural gaps identified through student feedback, the study aims to promote student engagement and reposition the course as a meaningful component of academic and professional development rather than a mere graduation requirement.

\subsection{Assessment Pattern in the Studied Courses}
The courses follow a continuous evaluation model, incorporating a combination of assignments, quizzes, projects, and examinations to assess student learning comprehensively. Examinations are conducted in both pen-and-paper and computer-based formats, while assignments are designed to foster both practical application and theoretical understanding. Quizzes, typically comprising multiple-choice questions, are administered before major exams to reinforce recently taught concepts and enhance retention \cite{articleMCQ}. Both multiple-choice and open-ended questions form key components of the assessment strategy and have proven effective in evaluating students’ programming knowledge. However, pen-and-paper exams have been criticized for their limited authenticity in assessing hands-on programming skills \cite{Bennedsen01102007}. To address this, the use of real-world assignments is increasingly recommended in computer science curricula \cite{articleL}. In the context of this study, project- and portfolio-based grading is also employed, where students compile a portfolio of work throughout the semester, which is evaluated at the end. This approach offers a holistic perspective on student progress and allows room for improvement, enabling recovery from occasional underperformance \cite{7062585}. Continuous evaluation extends to lab sessions as well, with weekly assessments and culminating in a comprehensive lab examination and viva that rigorously test their problem-solving skills. A persistent challenge for instructors is designing assignments that fall within students’ Zone of Proximal Development (ZPD), defined as the distance between the {actual developmental level} determined by independent problem solving and the {level of potential development} achievable through guided support or collaboration with more capable peers \cite{ZPD_1}.

\subsection{Role of TAs}
Teaching assistants (TAs) play a crucial role in handling increasing enrollment in CS courses by leading tutorials, conducting office hours, supervising lab sessions, and assisting with grading tasks \cite{TAs1}. Typically, master's or PhD students serve as TAs; however, many receive minimal formal training, which can pose significant challenges, particularly in assessment-related responsibilities \cite{10.1145/3291279.3339418}. TAs often face difficulties in addressing the diverse abilities and expectations of students in large, heterogeneous classrooms \cite{TAs1}. While collaborative grading among TAs, through peer calibration, has been shown to improve grading reliability, students may still perceive inconsistencies in evaluation practices \cite{TA_gradning}.

\subsection{Student Success Factors}
Student comfort and willingness to seek help are strong predictors of academic success in CS courses. Anxiety in CS education stems not only from programming challenges but also from test-related stress and collaborative tasks like pair programming~\cite{10.1145/2999541.2999557}. The CS classroom is often perceived as evaluative and competitive rather than supportive~\cite{article}. Persistent stereotypes portraying CS as male-dominated can hinder students’ sense of belonging, particularly among female students. Males often enter with prior CS experience, contributing to higher achievement, while females often report lower confidence, limited exposure, and a reduced sense of belonging, leading to lower grades \cite{female_3}. Many students prefer seeking help from peers with stronger programming skills rather than TAs, indicating greater trust in peer networks~\cite{10.1145/2729094.2742611}. To foster an inclusive environment, institutions must counter stereotypes, encourage collaboration, connect content to student interests, and promote engagement with TAs and instructors~\cite{inproceedingsF}. Confidence, prior experience, digital literacy, and sense of belonging significantly influence retention and can help reduce dropout rates in introductory CS courses~\cite{10.1145/2729094.2742611}.

\section{Method}
\label{sec:method}


\subsection{Study Framework}
This study builds on prior research conducted in a European educational setting and employs a comprehensive research design to explore key phenomena while validating the generalizability of previous findings in an Indian context. It began with qualitative data collection \cite{qualitative_step}, followed by quantitative data collection, preprocessing, and non-parametric statistical analysis to uncover underlying patterns \cite{10036401}. These insights informed the final structured quantitative phase presented in this study, which aims to validate and compare earlier findings within a different educational and cultural setting. Feedback statements adapted from \cite{10036401}, as listed in Table \ref{tab:themes_statements}, are analyzed across factors such as gender, academic branch, grades, and prior experience (refer to Table \ref{tab:pvalues_last_week_1}).

\subsection{Research Settings}
The study was conducted with first-year B.Tech. students from the 2022–2026 batch at IIT Patna during Spring 2023. All first-year students were required to enroll in an introductory programming and data structures course (CS102) and its associated laboratory component (CS112). Of the 516 enrolled students, 318 consistently participated in the weekly feedback surveys. A summary of participant information is presented in Table \ref{tab:metadata_analysis}. 
The course curriculum focused on C programming, covering topics such as basic syntax, data types, control flow, functions, pointers, fundamental data structures (e.g., linked lists, stacks, queues), file handling, and error management. The course was coordinated by three faculty members, resulting in a faculty-to-student ratio of 1:172, with additional support from TAs at a ratio of about 1:20. 
The assessment structure of CS102 included two assignments, two quizzes, a midterm, and an end-term examination, comprising multiple-choice, fill-in-the-blank, and justification-based questions. 
Weekly 3-hour lab sessions were evaluated through hands-on tasks, and midterm/ end-term lab exams of CS112 additionally included a viva component. 
A group project was assigned early in the semester and assessed near the end, just before the end-term exam. Students formed self-selected groups of three to five for collaborative tasks such as pair programming~\cite{4271597}. For project evaluation, they demonstrated running code and gave a brief presentation with Q\&A. 
While final grading was the responsibility of course instructors, TAs evaluated lab work, assignments, and quizzes \cite{10.1145/3488042.3488052}. This structure was designed to build foundational programming skills within a supportive and collaborative learning environment.


\begin{table}[]
\centering
\caption{Basic information of survey participants}
\begin{adjustbox}{width=1\linewidth}
\begin{tabular}{c|c|cc|cc|cc|cc|cc|cc|cc|cc|cc|ccc}
\hline
\multicolumn{2}{c|}{\textbf{Branch (B)}} & \multicolumn{2}{c|}{\textbf{AI}} & \multicolumn{2}{c|}{\textbf{CB}} & \multicolumn{2}{c|}{\textbf{CE}} & \multicolumn{2}{c|}{\textbf{CSE}} & \multicolumn{2}{c|}{\textbf{EE}} & \multicolumn{2}{c|}{\textbf{MC}} & \multicolumn{2}{c|}{\textbf{ME}} & \multicolumn{2}{c|}{\textbf{MM}} & \multicolumn{2}{c|}{\textbf{PH}} & \multicolumn{3}{c}{\textbf{Total}} \\ \hline

\multicolumn{2}{c|}{\textbf{Gender (G)}} & \textbf{M} & \textbf{F} & \textbf{M} & \textbf{F} & \textbf{M} & \textbf{F} & \textbf{M} & \textbf{F} & \textbf{M} & \textbf{F} & \textbf{M} & \textbf{F} & \textbf{M} & \textbf{F} & \textbf{M} & \textbf{F} & \textbf{M} & \textbf{F} & \textbf{M} & \textbf{F} & \textbf{M + F} \\ \hline 

\multicolumn{2}{c|}{\textbf{Student count}} & 28 & 6 & 30 & 10 & 25 & 11 & 40 & 12 & 35 & 7 & 19 & 5 & 39 & 10 & 19 & 3 & 16 & 3 & 251 & 67 & 318 \\ \hline 

\textbf{} & Nil & 19 & 5 & 25 & 7 & 22 & 11 & 31 & 9 & 28 & 7 & 16 & 5 & 32 & 6 & 14 & 3 & 9 & 3 & 196 & 56 & 252 \\ 
\textbf{PCE} & Medium & 9 & 1 & 5 & 3 & 2 & 0 & 9 & 3 & 7 & 0 & 3 & 0 & 7 & 3 & 5 & 0 & 6 & 0 & 53 & 10 & 63 \\ 
\textbf{} & High & 0 & 0 & 0 & 0 & 1 & 0 & 0 & 0 & 0 & 0 & 0 & 0 & 0 & 1 & 0 & 0 & 1 & 0 & 2 & 1 & 3 \\ \hline 

\textbf{} & Nil & 9 & 4 & 22 & 3 & 13 & 8 & 12 & 5 & 15 & 5 & 9 & 2 & 13 & 1 & 6 & 2 & 7 & 3 & 106 & 33 & 139 \\ 
\textbf{PAE} & Medium & 19 & 2 & 8 & 6 & 11 & 3 & 21 & 6 & 20 & 2 & 9 & 3 & 23 & 9 & 13 & 1 & 8 & 0 & 132 & 32 & 164 \\ 
\textbf{} & High & 0 & 0 & 0 & 1 & 1 & 0 & 7 & 1 & 0 & 0 & 1 & 0 & 3 & 0 & 0 & 0 & 1 & 0 & 13 & 2 & 15 \\ \hline 

\textbf{} & Nil & 2 & 3 & 6 & 1 & 3 & 1 & 2 & 1 & 4 & 1 & 3 & 0 & 4 & 0 & 1 & 0 & 0 & 0 & 25 & 7 & 32 \\ 
\textbf{PCoE} & Medium & 14 & 2 & 10 & 6 & 16 & 10 & 25 & 8 & 20 & 4 & 11 & 5 & 25 & 5 & 10 & 3 & 8 & 2 & 139 & 45 & 184 \\ 
\textbf{} & High & 12 & 1 & 14 & 3 & 6 & 0 & 13 & 3 & 11 & 2 & 5 & 0 & 10 & 5 & 8 & 0 & 8 & 1 & 87 & 15 & 102 \\ \hline 

\textbf{} & Nil & 0 & 0 & 0 & 0 & 1 & 0 & 0 & 0 & 0 & 0 & 0 & 0 & 3 & 0 & 0 & 0 & 0 & 0 & 4 & 0 & 4 \\ 
\textbf{PSE} & Medium & 12 & 1 & 5 & 6 & 9 & 8 & 13 & 4 & 12 & 5 & 4 & 2 & 15 & 3 & 1 & 1 & 0 & 1 & 71 & 31 & 102 \\ 
\textbf{} & High & 16 & 5 & 25 & 4 & 15 & 3 & 27 & 8 & 23 & 2 & 15 & 3 & 21 & 7 & 18 & 2 & 16 & 2 & 176 & 36 & 212 \\ \hline 

& AA & 9 & 2 & 1 & 2 & 0 & 1 & 17 & 2 & 9 & 0 & 9 & 2 & 11 & 2 & 6 & 0 & 3 & 0 & 65 & 11 & 76 \\ 
& AB & 7 & 1 & 3 & 0 & 6 & 2 & 11 & 4 & 10 & 3 & 8 & 0 & 15 & 3 & 3 & 0 & 3 & 0 & 66 & 13 & 79 \\ 
\textbf{Gd} & BB & 8 & 2 & 10 & 3 & 12 & 4 & 9 & 4 & 13 & 2 & 1 & 3 & 10 & 4 & 5 & 2 & 6 & 1 & 74 & 25 & 99 \\ 
& BC & 0 & 1 & 12 & 4 & 4 & 3 & 1 & 0 & 2 & 0 & 1 & 0 & 1 & 1 & 4 & 0 & 4 & 0 & 29 & 9 & 38 \\ 
& CC & 3 & 0 & 3 & 1 & 2 & 1 & 2 & 2 & 1 & 0 & 0 & 0 & 2 & 0 & 1 & 0 & 0 & 2 & 14 & 6 & 20 \\ 
& CD & 1 & 0 & 1 & 0 & 1 & 0 & 0 & 0 & 0 & 2 & 0 & 0 & 0 & 0 & 0 & 1 & 0 & 0 & 3 & 3 & 6 \\ \hline
\multicolumn{23}{r}{\textbf{PCE}: Prior C programming experience, 
\textbf{PAE}: Prior any programming experience, 
\textbf{PCoE}: Prior computer usage experience,} \\
\multicolumn{23}{r}{\textbf{PSE}: Prior smartphone usage experience, 
\textbf{AI}: Artificial Intelligence, 
\textbf{CB}: Chemical Engineering, 
\textbf{CE}: Civil Engineering,} \\
\multicolumn{23}{r}{\textbf{CSE}: Computer Science \& Engineering, 
\textbf{EE}: Electrical \& Electronics Engineering, 
\textbf{MC}: Mathematics \& Computing, } \\
\multicolumn{23}{r}{\textbf{ME}: Mechanical Engineering, 
\textbf{MM}: Metallurgy Engineering, 
\textbf{PH}: Engineering Physics,  
\textbf{Gd}: Grade, 
\textbf{M}: Male,
\textbf{F}: Female}

\end{tabular}
\end{adjustbox}
\label{tab:metadata_analysis}
\end{table}

\subsection{Data Collection}
An online portal was developed to administer the survey, comprising 25 mandatory statements (S1--S25) derived from previously established themes~\cite{10036401}, as listed in Table~\ref{tab:themes_statements}. 
To ensure respondent authenticity, the portal required login through the institute-issued official email ID. 
Students rated each statement on a 7-point Likert scale, where 1 indicated \emph{strongly disagree}, 7 indicated \emph{strongly agree}, and 4 represented a \emph{neutral} response. 
Survey links were distributed via email to all enrolled students at the end of each week over a span of nine weeks, following the completion of that week’s course activities. Participation was further encouraged through in-class announcements and follow-up reminders. 
The survey was conducted in English, which also served as the medium of instruction for the course. 
A total of 318 students voluntarily participated in the survey. At the beginning of the course, demographic and experiential background data were collected, including prior experience in C programming (PCE), general programming (PAE), computer usage (PCoE), and smartphone usage (PSE), each rated on a three-level ordinal scale: \textit{nil}, \textit{medium}, and \textit{high}. Gender and academic branch information were also recorded (see Table~\ref{tab:metadata_analysis}). 
Participation was completely voluntary and uncompensated. Students were informed of the study objectives, and informed consent was obtained. Ethical clearance was granted by the Institute Ethics Committee at IIT Patna (Ref: ITP/EC/2022-24/09, dated 10 July 2023).

\subsection{Data Analysis}
To address the research questions, the study employed a combination of descriptive and inferential statistical methods. Students were stratified into subgroups based on demographic and experiential variables, including gender, academic branch, course grade, and self-reported prior experience in C programming (PCE), general programming (PAE), computer use (PCoE), and smartphone use (PSE). 
The distribution of responses was first examined using Q-Q plots, which revealed that the data did not follow a normal distribution. 
Appendix \ref{App:qq_plots} 
provides the Q-Q plots for all 25 statements collected in Week-9. 
Given the non-normality of the data, the Mann-Whitney U test, a non-parametric test suitable for comparing two independent groups using ranked ordinal data, was applied to analyze gender-based differences. For comparisons across multiple subgroups (e.g., academic branches or prior experience levels), the Kruskal-Wallis test was employed, which is appropriate for assessing differences among three or more independent groups. 
Although course grades (AA–CD) are inherently ordinal, they were treated as continuous variables for the purpose of computing grade point averages (GPAs). All analyses were performed using the \emph{scipy.stats 1.16.1} library in Python. Statistical significance was evaluated at the 95\% confidence level across subgroups defined by 
academic branch (9 groups), 
gender (2 groups), 
grade (6 groups), and 
prior experience (3 levels), enabling a comprehensive understanding of response patterns and subgroup variations within the dataset.

\begin{table}[!hbt]
\centering
\caption{Survey statements}
\scriptsize 
\begin{adjustbox}{width=0.48\textwidth} 
\begin{tabular}{c |c p{7cm}} 
\hline
\textbf{Theme} & \textbf{SN.} & \textbf{Statement} \\
\hline
\multirow{8}{*}{\rotatebox{90}{\shortstack{Lab\\assignments}}} 
& S1: & I am experiencing the lab assignments foremost as an activity for me to learn the course content. \\
& S2: & I am experiencing the lab assignments as a necessary evil. \\
& S3: & I found it stressful to complete the lab assignments to the deadline. \\
& S4: & I have gotten a lot of help from my classmates when I solved the lab assignments. \\
& S5: & I have often helped other classmates when they solved the lab assignments. \\
\hline
\multirow{8}{*}{\rotatebox{90}{Examination}}
& S6: & I experienced that I knew in advance what types of questions would be on the exam. \\
& S7: & I experienced that I knew in advance what would be assessed on the written exam. \\
& S8: & I experienced that the type of knowledge and skills that were assessed on the written exam, are important to know/master when programming. \\
& S9: & I experienced the written exam as authentic, meaning that it consisted of questions I could encounter when developing programs. \\
\hline
\multirow{8}{*}{\rotatebox{90}{Project}} 
& S10: & I experienced that there was a big leap in difficulty between the lab assignments and the project. \\
& S11: & I experienced that I learnt a lot from doing the project assignment. \\
& S12: & I experienced the project assignments as authentic, meaning that they consisted of questions I could encounter when developing programs. \\
& S13: & I experienced it to be a huge difference between different project assignments when it comes to the amount of work that was required to achieve the same grade. \\
\hline
\multirow{6}{*}{\rotatebox{90}{\shortstack{Course\\Coordinator}}} 
& S14: & I experienced that the course coordinator has been able to put their own touch on the course, control the course structure and learning activities. \\
& S15: & I experienced that the course coordinator has had too little insight into, and shown too little interest in other course elements than lectures, such as tutorials and assessment of lab assignments. \\
\hline  
\multirow{19}{*}{\rotatebox{90}{Teaching Assistants (TAs)}} 
& S16: & I experienced that assessment of the lab assignments and projects is mainly done by the TAs. \\
& S17: & I experienced that there was a huge difference between how different TAs assessed the lab assignments during this course. \\
& S18: & I experienced that there was a huge difference between how different TAs assessed the project assignments during this course. \\
& S19: & I experienced that there was a huge difference between how much help you received from different TAs. \\
& S20: & I experienced that the structure and content of the tutorials differed depending on which TA you were assigned/chose to go to. \\
& S21: & I felt that the TAs in this course treated me professionally. \\
& S22: & I experienced that I got useful feedback and guidance from the TAs during the course. \\
& S23: & I felt it was easier to ask a TA for help, rather than the course coordinator. \\
& S24: & I preferred to search for answers on the internet, rather than asking a TA for help. \\
& S25: & I experienced that there often were too few TAs during the lab sessions, which made it difficult to get help or present my solution on time. \\
\hline
\end{tabular}
\end{adjustbox}
\label{tab:themes_statements}
\end{table}

\begin{figure}
    \centering
\begin{tikzpicture}
\begin{axis}[
    xbar stacked,
    bar width=6pt,
    width=0.48\textwidth,
    height=12cm,
    enlargelimits=0.05,
    xmin=-100, xmax=100,
    xmajorgrids = true,                
    grid style = {dashed, black!20},
    xtick={-100, -75, -50, -25, 0, 25, 50, 75, 100},
    xlabel={Likert scale response (\%)},
    symbolic y coords={S1,S2,S3,S4,S5,S6,S7,S8,S9,S10,S11,S12,S13,S14,S15,S16,S17,S18,S19,S20,S21,S22,S23,S24,S25},
    ytick=data,
    y dir=reverse,
    xticklabel style={/pgf/number format/fixed},
    tick label style={font=\footnotesize},
    label style={font=\footnotesize},
    legend style={at={(0.5,-0.1)}, anchor=north,legend columns=3, draw=none}
]

\addplot+[xbar, fill=gray!60, draw=black!100] coordinates {
(-20.91,S1) (-20.12,S2) (-21.22,S3) (-21.22,S4) (-22.01,S5)
(-23.11,S6) (-24.21,S7) (-22.48,S8) (-21.54,S9) (-26.10,S10)
(-16.66,S11) (-20.75,S12) (-25.94,S13) (-23.89,S14) (-21.22,S15)
(-25,S16) (-24.52,S17) (-28.30,S18) (-25,S19) (-26.10,S20)
(-22.95,S21) (-21.06,S22) (-24.68,S23) (-24.68,S24) (-25.94,S25)
};

\addplot+[xbar, fill=red!20, draw=black!100] coordinates {
(-4.40,S1) (-12.57,S2) (-12.89,S3) (-5.03,S4) (-7.86,S5)
(-11.32,S6) (-11.32,S7) (-4.40,S8) (-5.97,S9) (-10.37,S10)
(-4.40,S11) (-4.40,S12) (-11.32,S13) (-2.51,S14) (-13.83,S15)
(-7.86,S16) (-11.01,S17) (-11.01,S18) (-12.26,S19) (-9.11,S20)
(-8.49,S21) (-5.66,S22) (-8.17,S23) (-10.06,S24) (-11.32,S25)
};

\addplot+[xbar, fill=red!50, draw=black!100] coordinates {
(-2.83,S1) (-13.52,S2) (-15.09,S3) (-6.60,S4) (-3.14,S5)
(-8.80,S6) (-8.80,S7) (-2.20,S8) (-2.20,S9) (-10.37,S10)
(-0.94,S11) (-2.20,S12) (-6.60,S13) (-1.25,S14) (-15.09,S15)
(-6.28,S16) (-4.71,S17) (-4.08,S18) (-7.54,S19) (-7.86,S20)
(-3.14,S21) (-2.83,S22) (-2.51,S23) (-9.11,S24) (-10.69,S25)
};

\addplot+[xbar, fill=darkred, draw=black!100] coordinates {
(-0.62,S1) (-13.20,S2) (-11.01,S3) (-5.03,S4) (-1.25,S5)
(-4.08,S6) (-3.77,S7) (-0.62,S8) (-1.25,S9) (-4.71,S10)
(-0.31,S11) (-0.31,S12) (-4.08,S13) (-0.62,S14) (-11.94,S15)
(-4.71,S16) (-5.34,S17) (-5.66,S18) (-4.08,S19) (-5.34,S20)
(-2.20,S21) (-1.25,S22) (-2.20,S23) (-8.49,S24) (-8.80,S25)
};

\addplot+[xbar, fill=gray!60, draw=black!100, forget plot] coordinates {
(20.91,S1) (20.12,S2) (21.22,S3) (21.22,S4) (22.01,S5)
(23.11,S6) (24.21,S7) (22.48,S8) (21.54,S9) (26.10,S10)
(16.66,S11) (20.75,S12) (25.94,S13) (23.89,S14) (21.22,S15)
(25,S16) (24.52,S17) (28.30,S18) (25,S19) (26.10,S20)
(22.95,S21) (21.06,S22) (24.68,S23) (24.68,S24) (25.94,S25)
};

\addplot+[xbar, fill=green!20, draw=black!100] coordinates {
(18.55,S1) (13.83,S2) (13.52,S3) (23.89,S4) (23.58,S5)
(21.06,S6) (17.61,S7) (20.75,S8) (21.06,S9) (16.35,S10)
(28.61,S11) (22.64,S12) (16.98,S13) (24.21,S14) (13.52,S15)
(14.15,S16) (17.92,S17) (12.26,S18) (15.40,S19) (15.09,S20)
(19.18,S21) (21.06,S22) (19.18,S23) (13.20,S24) (11.63,S25)
};

\addplot+[xbar, fill=green!50, draw=black!100] coordinates {
(15.09,S1) (5.03,S2) (3.45,S3) (11.32,S4) (12.89,S5)
(5.03,S6) (6.60,S7) (17.29,S8) (15.40,S9) (4.08,S10)
(20.44,S11) (18.86,S12) (6.28,S13) (14.77,S14) (2.51,S15)
(11.32,S16) (6.60,S17) (5.66,S18) (7.54,S19) (7.23,S20)
(12.57,S21) (16.03,S22) (12.26,S23) (4.40,S24) (3.45,S25)
};

\addplot+[xbar, fill=darkgreen, draw=black!100] coordinates {
(16.66,S1) (1.57,S2) (1.57,S3) (5.66,S4) (7.23,S5)
(3.45,S6) (3.45,S7) (9.74,S8) (11.01,S9) (1.88,S10)
(11.94,S11) (10.01,S12) (2.83,S13) (8.80,S14) (0.62,S15)
(5.66,S16) (5.34,S17) (4.74,S18) (3.14,S19) (3.14,S20)
(8.49,S21) (11.01,S22) (6.28,S23) (5.34,S24) (2.20,S25)
};


\addlegendimage{xbar, fill=gray!60, draw=none}
\addlegendentry{\textcolor{gray!100}{neutral}}

\addlegendimage{xbar, fill=red!20, draw=none}
\addlegendentry{\textcolor{red!80}{weakly disagree}}

\addlegendimage{xbar, fill=red!50, draw=none}
\addlegendentry{\textcolor{red!90}{disagree}}

\addlegendimage{xbar, fill=darkred, draw=none}
\addlegendentry{\textcolor{darkred}{strongly disagree}}

\addlegendimage{xbar, fill=green!20, draw=none}
\addlegendentry{\textcolor{green!90}{weakly agree}}

\addlegendimage{xbar, fill=green!50, draw=none}
\addlegendentry{\textcolor{green!100}{agree}}

\addlegendimage{xbar, fill=darkgreen, draw=none}
\addlegendentry{\textcolor{darkgreen}{strongly agree}}

\end{axis}

\end{tikzpicture}

\caption{Distribution of Likert-scale responses for all 25 survey statements, based on feedback from all 318 students}
\label{fig:likert_plot}

\end{figure}
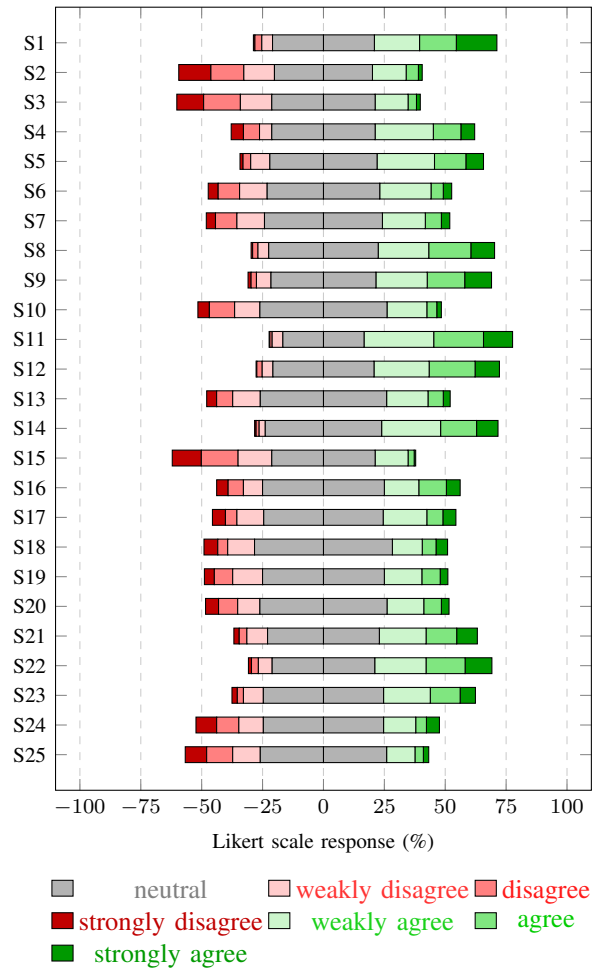

\section{Results} \label{4sec:result}
This section presents results derived from the survey data and their subsequent analysis. Prior experience variables were encoded on an ordinal scale: \emph{nil} = 0, \emph{medium} = 1, and \emph{high} = 2. Academic grades were mapped to numerical values as follows: AA = 10, AB = 9, BB = 8, BC = 7, CC = 6, and CD = 5, with AA representing the highest possible grade on a 10-point scale. 
While the survey spanned nine weeks, the main paper focuses on Week-9 responses, which offer the most comprehensive feedback following the completion of course activities. Results from earlier weeks are included in the supplementary material.

\subsection{Student Perceptions of Assessment}
The survey findings related to lab assignments (S1–S5), exams (S6–S9), and projects (S10–S13) are presented in the top half of Fig.~\ref{fig:likert_plot}. While many students selected neutral responses across these items, a substantial number perceived lab assignments as meaningful learning activities rather than obligatory tasks. These assignments were generally not seen as stressful, with several students citing peer collaboration as a key factor in managing deadlines.  
Student responses regarding awareness of exam content and question types were varied, with no clear consensus. However, many acknowledged that the skills and knowledge assessed in the exams were both important and beneficial, and viewed the exams as authentic assessments. 
Regarding projects, a significant proportion of students remained neutral on whether there was a notable increase in difficulty compared to lab assignments. Nevertheless, some students noted variations in the workload required to attain similar grades across different project topics. Despite this, the majority found the projects to be authentic learning experiences and reported substantial knowledge gains from completing them.

\subsection{Student Perceptions of Course Coordinators and TAs}
The lower part of Fig.~\ref{fig:likert_plot} shows that for statements S14 and S15, which relate to the course coordinator's effectiveness, students generally acknowledged the coordinator’s active engagement with various course components beyond just lectures. Overall, students perceived the coordinator as well-informed and interested in learning activities.

Regarding TAs (statements S16–S25 in Fig.~\ref{fig:likert_plot}), student responses showed considerable variation in perceptions of TA involvement in assessing lab assignments and projects. Many students reported being treated professionally by TAs, appreciated the feedback and guidance they received, and found it easier to seek assistance from TAs than from course coordinators. 
However, most students remained neutral regarding inconsistencies in how different TAs assessed lab and project work, the level of help provided, and the content delivered during tutorials. Interestingly, and contrary to conventional expectations, a notable proportion of students preferred consulting TAs over searching for solutions online. While opinions on TA availability varied, a significant number of students indicated that adequate teaching support was available during course activities.


\begin{table}[!ht]

    \caption{Subgroup comparison of $p$-values using Kruskal-Wallis test/ Mann-Whitney U test}
    \centering
    \begin{adjustbox}{width=0.4\textwidth} 
    \begin{tabular}{l| cccc ccc}
\cline{2-8}
\multicolumn{1}{c}{}         & \textbf{G} & \textbf{PCE} & \textbf{PAE} & \textbf{PCoE} & \textbf{PSE} & \textbf{Gd} & \textbf{B} \\
        \hline 

        \textbf{G} & 0 & 0.506 & 0.32 & 0.153 & \textbf{0.049} & \textbf{0.017} & 0.782 \\

        \textbf{PCE} & 0.475 & 0 & \textbf{<0.001} & \textbf{<0.001} & \textbf{<0.001} & 0.448 & 0.217 \\

        \textbf{PAE} & 0.504 & \textbf{<0.001} & 0 & \textbf{<0.001} & \textbf{<0.001} & \textbf{<0.001} & \textbf{0.015} \\

        \textbf{PCoE} & 0.153 & \textbf{<0.001} & \textbf{<0.001} & 0 & \textbf{<0.001} & \textbf{0.051} & 0.993 \\

        \textbf{PSE} & \textbf{0.014} & \textbf{0.002} & \textbf{<0.001} & \textbf{<0.001} & 0 & 0.658 & 0.361 \\

        \textbf{Gd} & 0.148 & 0.762 & \textbf{<0.001} & 0.096 & 0.057 & 0 & \textbf{<0.001} \\

        \textbf{B} & 0.836 & 0.296 & \textbf{0.01} & 0.398 & \textbf{0.004} & \textbf{0.001} & 0 \\
\hline 
    \end{tabular}
    \end{adjustbox}
    \label{tab:meta_data_pvalues}
\end{table}

\subsection{Differences}

Table~\ref{tab:meta_data_pvalues} summarizes the results of statistical tests examining differences across various subgroups, including 
gender (G), academic performance grade (Gd), branch (B), and 
prior experiences in C programming (PCE)/ any programming language (PAE)/ computer usage (PCoE)/ smartphone usage (PSE). 
Key observations are discussed below.

\subsubsection{\textbf{Gender (G)}}
In this study, female students were underrepresented compared to their male counterparts (see Table~\ref{tab:metadata_analysis}). As shown in Table~\ref{tab:meta_data_pvalues}, a statistically significant difference was observed in PSE (Mann-Whitney U test, $p$-value = 0.049), with male students reporting slightly higher engagement level. This may reflect a tendency for greater male participation in technology-oriented activities. 
Additionally, female students achieved lower average grades than male students (Mann-Whitney U test, $p$-value = 0.017), with a mean cumulative GPA of 8.07 compared to 8.52 for males.



\subsubsection{\textbf{Prior C Programming Experience (PCE)}}
As shown in Table~\ref{tab:meta_data_pvalues}, significant differences were observed among student groups based on PCE. The Kruskal-Wallis test indicated strong disparities in PAE across PCE levels ($p$~<~0.001). Post hoc Dunn–Bonferroni pairwise comparisons further revealed that students with medium PCE significantly differed from those with no PCE (adjusted $p$~<~0.001). Similar significant variations were found in PCoE and PSE, with $p$-values less than 0.001 for both. 
These results suggest that PCE is strongly associated with broader technological familiarity. Specifically, students with PCE also tended to report higher levels of programming experience in general (PAE), as well as more frequent and confident use of computers and smartphones (PCoE and PSE). 
These findings suggest that early exposure to programming, particularly in C, supports the development of coding skills through enhanced logical thinking. Students with PCE also tend to engage more with digital tools, likely contributing to greater technical proficiency across platforms.

\subsubsection{\textbf{Prior Any programming Experience (PAE)}}
A substantial difference in academic performance was observed when students were grouped by PAE, as indicated by the Kruskal-Wallis test ($p$~<~0.001). Students with no PAE had lower average grades (mean = 7.91) compared to those with moderate (mean = 8.79) and high (mean = 9.13) PAE. This suggests a strong positive association between PAE and academic achievement (grade), indicating that foundational programming skills contribute to improved learning outcomes in subsequent courses. 
In addition, academic branch distribution varied significantly across PAE groups (Kruskal-Wallis test, $p$ = 0.015). Post hoc Dunn-Bonferroni comparisons revealed a statistically significant difference between students with no PAE and those with high PAE (adjusted $p$ = 0.024). 
%
These findings underscore the importance of prior programming experience in shaping both academic success and disciplinary pathways among engineering students.


\subsubsection{\textbf{Prior Computer usage Experience (PCoE)}}
A near-significant difference in grades was observed across PCoE groups (Kruskal–Wallis test, $p$ = 0.051). Students with no prior computer experience had a lower average grade (mean = 7.91) compared to those with moderate (mean = 8.44) and high (mean = 8.56) PCoE. Post hoc analysis showed a statistically significant difference between students with no and high PCoE (adjusted $p$ = 0.047).
These findings suggest a positive association between computer familiarity and academic performance; students with greater experience using computers tend to perform better in the course.


\subsubsection{\textbf{Prior Smartphone usage Experience (PSE)}}
Grouping students by PSE revealed a significant gender-related difference (Kruskal-Wallis test, $p$ = 0.014). Notably, all students with no reported PSE were male. A statistically significant difference was also observed between moderate and high PSE levels (adjusted $p$ = 0.019). As shown in Table~\ref{tab:metadata_analysis}, 70.11\% of male students reported high smartphone usage, compared to 53.73\% of female students. 
This suggests a potential gender-related pattern in reported smartphone usage, possibly reflecting differences in technology adoption trends influenced by societal norms or personal preferences.


\subsubsection{\textbf{Grade (Gd)}} 
A significant difference in PAE was observed when students were grouped by academic grade (Kruskal-Wallis test, $p$~<~0.001), with multiple grade pairs showing statistically significant differences: 
AA–CD (adjusted $p$ = 0.004), 
AA–CC (adjusted $p$ = 0.001), 
BC–AA (adjusted $p$ < 0.001), 
BB–AA (adjusted $p$ < 0.001), 
AB–CD (adjusted $p$ = 0.024), 
AB–CC (adjusted $p$ = 0.017), and 
BC–AB (adjusted $p$ = 0.026). 
Students with higher grades, AA (mean PAE = 0.87) and AB (mean PAE = 0.75), tended to report more PAE. In contrast, students with lower grades, particularly those receiving CD (mean PAE = 0), CC (mean PAE = 0.3), and BC (mean PAE = 0.39), had less PAE. 
This suggests that foundational programming exposure may contribute to improved performance in subsequent technical coursework.


\subsubsection{\textbf{Branch (B)}}
CSE students reported the highest prior programming experience (PAE) with a mean of 0.83, while the CB students had the lowest (mean PAE = 0.40), as comprehended from Table~\ref{tab:metadata_analysis}. CE students reported the lowest PSE, with a mean of 1.47. 
Statistically significant differences in academic grades were observed across branches (Kruskal-Wallis test, $p$ < 0.001). Post hoc pairwise comparisons revealed notable differences: 
CB vs. CSE (adjusted $p$ < 0.001), 
CE vs. CSE (adjusted $p$ < 0.001), 
CB vs. MC (adjusted $p$ < 0.001), and 
CE vs. MC (adjusted $p$ < 0.001). 
Students in the CB and CE branches had lower average grades (both < 8.0), whereas students in MC (mean = 9.20) and CSE (mean = 8.84) performed better. Other branches fell in between, with average grades ranging from 8.0 to 8.8. 
These results suggest that academic performance and prior technical experience vary notably by discipline, with students from computing-focused branches (CSE, MC) generally exhibiting stronger preparation and outcomes.


Table~\ref{tab:pvalues_last_week_1} provides a detailed Week-9 analysis; while results from initial eight weeks are available in 
\ref{App:weeks_analysis}.


\begin{table*}[]



\caption{\footnotesize{Week-9 analysis on $p$-values across subgroups: Mann-Whitney U test for gender (G), and Kruskal-Wallis test based on academic grade (Gd), branch (B), and prior experiences in C programming (PCE)/ any programming language (PAE)/ computer usage (PCoE)/ smartphone usage (PSE)}}

\begin{adjustbox}{rotate=90, width=1\textwidth}
\begin{tabular}{c|lp{2.1cm}|ccccccc|p{20cm}}

\hline
\textbf{Theme} & \textbf{SN.} & \textbf{Statement} & \textbf{G} & \textbf{PCE} & \textbf{PAE} & \textbf{PCoE} & \textbf{PSE} & \textbf{Gd} & \textbf{B} & \multicolumn{1}{c}{\textbf{Post-hoc test results explanations}}   \\ \hline \hline

\multirow{12}{*}{\rotatebox{270}{\shortstack{Lab\\Assignments}}} & S1: & Learning activity & 0.903 & 0.582 & \textcolor{darkred}{ \textbf{0.018}} & 0.313 & \textcolor{darkred}{\textbf{0.048}} & 0.400 & 0.763 & \textcolor{darkred}{\textbf{PAE}}: Students with \emph{high} PAE disagreed to a greater extent, significantly differing from those with \emph{medium} PAE (adj. $p$ =  0.047). 
\textcolor{darkred}{\textbf{PSE}}: Students with \emph{nil} PSE disagreed to a greater extent, significantly differing from those with \emph{high} (adj. $p$ =  0.041) and \emph{medium} PSE (adj. $p$ =  0.041).  \\

& S2: & Necessary evil & 0.968 & \textcolor{darkgreen}{\textbf{0.154}} & 0.242 & 0.431 & 0.469 & 0.697 & \textcolor{darkgreen}{\textbf{0.139}} & \textcolor{darkgreen}{\textbf{B}}: CB students (mean =  3.95) and EE students (mean =  3.83) showed the highest level of agreement compared to other branches, while PH students disagreed the most, with a mean of 2.95.  \\

& S3: & Stressful & 0.486 & 0.425 & 0.483 & 0.276 & 0.382 & \textcolor{darkgreen}{\textbf{0.107}} & \textcolor{darkgreen}{\textbf{0.144}} & \textcolor{darkgreen}{\textbf{B}}: CB students showed the highest agreement (mean =  4.03), while PH students had the highest level of disagreement (mean =  3.11). \textcolor{darkgreen}{\textbf{Gd}}: Students with a BC grade exhibited the highest agreement (mean =  3.79), whereas those with a CD grade showed the most disagreement (mean =  2.95).  \\

& S4: & \shortstack{Help from peers} & 0.469 & 0.400 & 0.791 & 0.320 & 0.308 & 0.234 & \textcolor{blue}{\textbf{0.094}} & \textcolor{blue}{\textbf{B}}: Students from different branches generally agreed, except for those in CSE and CE, who expressed disagreement.  \\

& S5: & Help to peers & 0.887 & \textcolor{darkgreen}{\textbf{0.175}} & \textcolor{darkgreen}{\textbf{0.158}} & 0.448 & 0.599 & 0.998 & 0.634 &  \textcolor{darkgreen}{\textbf{PCE}}: Students with \emph{nil} (mean =  4.58) and \emph{medium} (mean =  4.40) PCE strongly agreed, whereas those with \emph{high} PCE disagreed the most (mean =  3.67). \textcolor{darkgreen}{\textbf{PAE}}: Students with \emph{high} PAE showed more disagreement (mean =  4.14) than \emph{medium} (mean =  4.59) and \emph{nil} (mean =  4.51) PAE.  \\ \hline

\multirow{9}{*}{\rotatebox{270}{\shortstack{Examination}}} & S6: & Predictable question types & 0.497 & \textcolor{darkgreen}{\textbf{0.117}} & 0.237 & \textcolor{blue}{\textbf{0.093}} & \textcolor{darkgreen}{\textbf{0.121}} & 0.591 & 0.418 &  \textcolor{blue}{\textbf{PCoE}}: Students with \emph{medium} PCoE (mean =  3.87) disagreed more strongly than those with a \emph{high} (mean =  4.17) or \emph{nil} (mean =  4.25) PCoE. \textcolor{darkgreen}{\textbf{PCE}}: Students with \emph{nil} PCE (mean =  3.94) disagreed more strongly than those with a lot (mean =  4.33) or \emph{medium} (mean =  4.24) PCE. \textcolor{darkgreen}{\textbf{PSE}}: Students with \emph{high} PSE (mean =  4.11) disagreed more strongly than those with \emph{medium} PSE (mean =  3.77). \\

&  S7: & Predictable assessment & 0.799 & 0.555 & 0.462 & \textcolor{darkgreen}{\textbf{0.179}} & 0.648 & 0.236 & 0.683 & \textcolor{darkgreen}{\textbf{PCoE}}: Students with \emph{nil} PCoE showed the highest level of agreement (mean =  4.34), followed by those with \emph{high} (mean =  4.14) and \emph{medium} PCoE (mean =  3.88).  \\

& S8: & Skills \& knowledge & 0.973 & 0.631 & \textcolor{darkgreen}{\textbf{0.141}} & 0.630 & \textcolor{darkgreen}{\textbf{0.114}} & \textcolor{darkgreen}{\textbf{0.170}} & 0.497 & \textcolor{darkgreen}{\textbf{PSE}}: Students with \emph{nil} PSE tended to report stronger disagreement (mean =  3.75) than those with \emph{medium} (mean =  4.83) and \emph{high} (mean =  4.71) PSE. \textcolor{darkgreen}{\textbf{Gd}}: Students with AA grade showed the strongest agreement (mean =  4.97), followed by those with BC (mean =  4.89), CC (mean =  4.90), CD (mean =  4.83), AB (mean =  4.68), and BB (mean =  4.51). \\

& S9: & Authentic & 0.462 & 0.977 & \textcolor{blue}{\textbf{0.075}} & 0.973 & 0.359 & 0.373 & 0.379 & \textcolor{blue}{\textbf{PAE}}: Moderate significance was observed.   \\ \hline

\multirow{7}{*}{\rotatebox{270}{\shortstack{Project}}} &  
S10: & Big leap & 0.282 & 0.339 & 0.628 & 0.368 & 0.502 & 0.618 & 0.415 & ---   \\

& S11: & Learnt a lot & 0.442 & 0.460 & \textcolor{darkred}{\textbf{0.017}} & 0.443 & \textcolor{blue}{\textbf{0.069}} & 0.263 & 0.824 & \textcolor{darkred}{\textbf{PAE}}: Students with \emph{medium} PAE showed the highest agreement (mean =  5.13), significantly differing from those with \emph{nil} PAE (mean =  4.84, adj. $p$ =  0.028). Students with \emph{high} PAE had the lowest agreement (mean =  4.67). \textcolor{blue}{\textbf{PSE}}: Students with \emph{nil} PSE showed the lowest agreement (mean =  3.5), significantly differing from those with \emph{medium} (mean =  5.1, adj. $p$ =  0.07) and \emph{high} PSE (mean =  4.95, adj. $p$ =  0.107). \\

& S12: & Authentic & 0.690 & 0.267 & \textcolor{darkred}{\textbf{0.006}} & 0.743 & 0.823 & 0.736 & 0.657 & \textcolor{darkred}{\textbf{PAE}}: Students with \emph{medium} prior programming experience showed the highest agreement (mean =  4.98), significantly differing from those with a lot of prior experience (mean =  4.13, adj. $p$ =  0.024) and those with \emph{nil} prior experience (mean =  4.68, adj. $p$ =  0.069).  \\

& S13: & Difference & 
0.604 & 0.229 & 0.606 & 0.520 & 0.727 & 0.433 & 0.873 & ---   \\ \hline

\multirow{6}{*}{\rotatebox{270}{\shortstack{Course\\Coordinator}}} & S14: & \small{Course structure} & 
0.568 & 0.402 & \textcolor{darkgreen}{\textbf{0.142}} & \textcolor{blue}{\textbf{0.082}} & 0.368 & 0.470 & 0.842 & \textcolor{darkgreen}{\textbf{PAE}}: Students with \emph{high} PAE showed stronger agreement. \textcolor{blue}{\textbf{PCoE}}: All students strongly agreed. \\

& S15: & Little insight & \textcolor{darkgreen}{\textbf{0.122}} & 0.701 & \textcolor{darkred}{\textbf{0.001}} & \textcolor{blue}{\textbf{0.074}} & 0.381 & 0.516 & 0.214 & \textcolor{darkgreen}{\textbf{G}}: The male group had equally \emph{high} values (median = 4) as the female group (median = 4). Male students showed \emph{high}er agreement (mean =  3.46) compared to female students (mean =  3.19). \textcolor{darkred}{\textbf{PAE}}: Pairwise group comparisons between \emph{nil} - \emph{medium} and \emph{high} - \emph{medium} showed adj. $p$-values less than 0.05. Students with \emph{medium} PAE (mean =  3.09) strongly disagreed, while students with \emph{high} (mean =  4.2) and \emph{nil} PAE (mean =  3.69). \textcolor{blue}{\textbf{PCoE}}: Students with \emph{high} PCoE (mean =  3.16) strongly disagreed, while students with \emph{medium} (mean =  3.49) and \emph{nil} PCoE (mean =  3.69). \\ \hline

\multirow{26}{*}{\rotatebox{270}{\shortstack{Teaching Assistants (TAs)}}} &  S16: & Assessment & 0.956 & \textcolor{darkgreen}{\textbf{0.133}} & \textcolor{darkgreen}{\textbf{0.158}} & 0.384 & 0.338 & 0.354 & 0.618 & \textcolor{darkgreen}{\textbf{PCE}}: Students with \emph{medium} PCE (mean =  3.89) somewhat disagreed compared to students with \emph{nil} PCE (mean =  4.26). \textcolor{darkgreen}{\textbf{PAE}}: All students were neutral in their response, with an average mean of 4.08. \\

&  S17: & Variation in lab assessment & 0.694 & \textcolor{darkgreen}{\textbf{0.119}} & 0.681 & \textcolor{blue}{\textbf{0.097}} & 0.261 & 0.876 & 0.337 & \textcolor{darkgreen}{\textbf{PCE}}: Students with \emph{medium} PCE (mean =  3.83) showed stronger disagreement compared to those with \emph{high} PCE (mean =  4.67). \textcolor{blue}{\textbf{PCoE}}: Overall, students were neutral in their responses, with an average mean of 4.21.  \\

& S18: & Variation for projects & 0.870 & \textcolor{blue}{\textbf{0.073}} & 0.710 & \textcolor{blue}{\textbf{0.068}} & \textcolor{darkgreen}{\textbf{0.141}} & 0.845 & \textcolor{darkgreen}{\textbf{0.174}} & \textcolor{blue}{\textbf{PCoE}}: Overall, students were neutral in their responses, with an average mean of 4.15. \textcolor{darkgreen}{\textbf{B}}: CB students showed stronger agreement (mean =  4.5), while AI (mean =  3.79), ME (mean =  3.78), and PH (mean =  3.68) students showed comparatively lower agreement.  \\

& S19: & Variation for help & 0.906 & 0.244 & 0.697 & 0.732 & \textcolor{darkgreen}{\textbf{0.158}} & 0.577 & 0.305 & \textcolor{darkgreen}{\textbf{PSE}}: Students with \emph{nil} PSE strongly disagreed (mean =  3.25), while those with \emph{medium} to \emph{high} PSE were generally neutral (mean =  3.99).  \\

& S20: & Tutorials & 0.739 & 0.532 & 0.593 & 0.371 & 0.834 & 0.873 & 0.817 & ---   \\

& S21: & Professionalism & 0.674 & 0.543 & \textcolor{darkred}{\textbf{0.018}} & 0.909 & \textcolor{darkgreen}{\textbf{0.141}} & 0.655 & 0.863 & \textcolor{darkred}{\textbf{PAE}}: All adj. $p$-values were greater than 0.05, indicating no significant pairwise differences. However, students with \emph{medium} PAE showed stronger agreement (mean =  4.66) compared to those with \emph{nil} or \emph{high} PAE (weighted mean =  4.29). \textcolor{darkgreen}{\textbf{PSE}}: Students with \emph{medium} (mean =  4.56) and \emph{high} PSE (mean =  4.47) strongly agreed, whereas those with \emph{nil} PSE strongly disagreed (mean =  3.5). \\

& S22: & Feedback & 0.650 & 0.913 & \textcolor{darkgreen}{\textbf{0.197}} & 0.922 & 0.947 & \textcolor{darkred}{ \textbf{0.010}} & 0.931 & \textcolor{darkgreen}{\textbf{PAE}}: All students strongly agreed, with a weighted mean of 4.71. \textcolor{darkred}{\textbf{Gd}}: A significant difference was observed between students with BC and BB grades (adj. $p$ =  0.009). Students with grades BC (mean =  5.24), CC (mean =  5.10), CD (mean =  4.83), and AA (mean =  4.83) showed strong agreement. In contrast, students with BB grade (mean =  4.37) showed comparatively lower agreement. \\

& S23: & Asking TA help & 0.505 & 0.411 & 0.894 & 0.744 & \textcolor{blue}{\textbf{0.100}} & 0.269 & 0.571 & \textcolor{blue}{\textbf{PSE}}: Students with \emph{medium} to \emph{high} PSE strongly agreed, with a weighted mean of 4.44. In contrast, students with \emph{nil} PSE strongly disagreed (mean =  3.5).  \\

& S24: & Preferred internet & 0.382 & 0.763 & \textcolor{blue}{\textbf{0.091}} & \textcolor{blue}{\textbf{0.091}} & \textcolor{darkred}{\textbf{0.027}} & 0.895 & 0.882 & \textcolor{blue}{\textbf{PAE}}: Students with \emph{high} PAE agreed (mean =  4.33), while those with \emph{nil} PAE disagreed (mean =  3.68). Students with \emph{medium} PAE were nearly neutral (mean =  3.94). \textcolor{blue}{\textbf{PCoE}}: Students with \emph{nil} (mean =  3.59) and \emph{medium} (mean =  3.73) PCoE disagreed, with a weighted mean of 3.70. In contrast, students with \emph{high} PCoE were somewhat neutral (mean =  4.12). \textcolor{darkred}{\textbf{PSE}}: A significant difference was observed between students with \emph{high} and \emph{medium} PSE (adj. $p$ =  0.025). Students with \emph{medium} PSE strongly disagreed (mean =  3.52), while those with \emph{high} PSE were more agreeable (mean =  4.00). \\

& S25: & Few TAs & 0.890 & \textcolor{darkred}{ \textbf{0.008}} & 0.572 & 0.488 & \textcolor{blue}{\textbf{0.098}} & 0.780 & 0.449 & \textcolor{darkred}{\textbf{PCE}}: Significant differences were found in pairwise comparisons between \emph{medium} - \emph{high} (adj. $p$ =  0.007) and \emph{high} - \emph{nil} (adj. $p$ =  0.014). Students with \emph{medium} (mean =  3.46) and \emph{nil} (mean =  3.68) PCE (weighted mean =  3.66) strongly disagreed, whereas students with \emph{high} PCE (mean =  6.00) strongly agreed. \textcolor{blue}{\textbf{PSE}}: However, students with \emph{high} (mean =  3.73) and \emph{medium} PSE (mean =  3.49, weighted mean =  3.65) strongly disagreed, while those with \emph{nil} PSE (mean =  4.50) were more agreeable. \\ \hline



\multicolumn{11}{r}{
\textcolor{darkred}{\textbf{Red}} $p$-values indicate strong significance ($p \leq 0.05$), 
\textcolor{blue}{\textbf{blue}} indicate moderate significance ($0.05 < p \leq 0.1$), and 
\textcolor{darkgreen}{\textbf{green}} denote marginal trends ($0.1 < p \leq 0.2$).
}

\label{tab:pvalues_last_week_1}
\end{tabular}
\end{adjustbox}
\end{table*}

\section{Discussion and Conclusion}
\label{5sec:discussion_conclusion}


This section interprets the results through theoretical perspectives, discusses practical implications, outlines the study’s limitations, and suggests directions for future research.

\subsection{Student Perceptions of Assessment}
The findings indicate that students perceived lab assignments (S1) as valuable learning activities (refer to Table \ref{tab:themes_statements} for details), aligning well with the principles of formative assessment \cite{theory_perspective}. They did not view them as a necessary evil (S2) \cite{qualitative_step}, suggesting that the assessments were seen as tools to enhance learning rather than merely to certify achievement. Students’ emphasis on collaborative efforts during labs resonates with social constructivist theories \cite{collab_perspective_1}, where peer interaction and shared problem-solving foster knowledge construction \cite{collab_perspective,student_opinion}. Lower stress levels associated with lab sessions (S3) may be attributed to constructive alignment \cite{constructive_alignment}, where coherence between learning objectives, activities, and assessments, combined with peer support (S4 and S5), helps reduce anxiety \cite{10.1145/2999541.2999557}.

Students showed varying levels of awareness about the exam content (S6 and S7), a trend also reflected in the qualitative findings \cite{qualitative_step}. However, many perceived exams as authentic and meaningful assessments of skills and knowledge (S8 and S9), aligning with the theory of authentic assessment \cite{authentic_assessment} and previous findings \cite{10036401}. Although the exams primarily relied on conventional formats such as multiple-choice questions, students’ perception of authenticity may stem from clear and transparent expectations, an essential factor in reducing assessment-related anxiety \cite{authentic_assessment_1}.

Projects were generally seen as authentic learning experiences (S12), consistent with experiential learning theory \cite{learning}, which emphasizes the educational value of hands-on engagement. Students also reported substantial learning gains from completing projects (S11). However, some noted inconsistencies in the effort required to achieve similar grades (S13), raising concerns about perceived fairness in assessment. Neutral responses regarding the increased difficulty from lab assignments to projects (S10) may suggest that the jump in complexity was too steep and possibly beyond some students' ZPD \cite{ZPD_1}. This highlights to the need for a more gradual increase in task difficulty to better scaffold student learning and support their progression.

\subsection{Student Perceptions of Course Coordinator and TAs}
The largely neutral responses of students to the effectiveness of the course coordinator, with a notable subset recognizing their active participation beyond lectures (S14), suggest that the coordinator was perceived as accessible and involved in multiple aspects of the course. This perception aligns with the concept of distributed leadership in higher education, where instructional responsibilities are shared across various roles to promote collective responsibility for student learning \cite{harris2022distributed}. In this model, leadership extends beyond direct instruction to include coordination of assessments, learning resources, and support mechanisms. The absence of negative perceptions about the coordinator’s insight into non-lecture activities (S15) may reflect the benefits of visible involvement and transparent communication, both of which foster student trust and satisfaction.

Regarding TAs (S16-S25), students generally reported positive experiences with professional behavior, constructive feedback, and approachability. Notably, students found TAs more accessible than the course coordinator, a finding consistent with prior research emphasizing the role of TA approachability and social presence in fostering help-seeking behavior and academic engagement \cite{wilson2002study, 10.1145/3291279.3339418}. Interestingly, many students indicated a preference for consulting TAs over online resources when encountering challenges. This shift may highlight a growing appreciation for personalized, context-aware support within the classroom, possibly enabled by the availability of a sufficient number of TAs (S25). This contrasts with earlier qualitative findings \cite{qualitative_step}, suggesting that improvements in TA accessibility may have occurred during the studied course. 
However, notable variability was observed in students' perceptions of TA consistency regarding assessment, assistance, and tutorial content. Such inconsistencies can negatively impact students’ perceptions of fairness and uniformity, aligning with concerns raised in prior studies that emphasize the need for standardized training and assessment guidelines for TAs \cite{TAs1, TAS2}. Although the majority of students acknowledged the presence of adequate TA support, the uneven perceptions of their availability suggest potential mismatches between support distribution and student needs during high-demand periods.

These findings underscore the importance of structured TA coordination to ensure consistent and equitable support. All students must receive timely guidance and formative feedback, especially for tasks situated within their ZPD, i.e., tasks they cannot yet complete independently but can accomplish with appropriate scaffolding.

\subsection{Differences between Subgroups of Students}
The observed differences in academic performance and prior experience across gender, programming exposure, and academic branches can be contextualized through multiple theoretical lenses. The relatively lower grades and limited prior programming experience reported by female students align with {stereotype threat theory} \cite{diff_theory_1}, which suggests that societal stereotypes about gender roles in STEM can undermine confidence and hinder performance. The gender gap in smartphone and computer usage, e.g., 70.11\% of males reporting high PSE compared to 53.73\% of females, may reflect patterns of {gender socialization} within the Indian context \cite{diff_theory_2}. This disparity in early exposure can affect {self-efficacy} \cite{diff_theory_3}, influencing both academic engagement and learning outcomes. As previous studies have shown \cite{inproceedingsF, female_3}, the under-representation of women in technical branches can further exacerbate these challenges. Isolation in male-dominated environments may reduce the sense of belonging \cite{diff_theory_4}, thereby negatively affecting performance.

The strong correlation among prior experiences (PCE, PAE, PCoE, PSE) and academic success supports the theory of {transfer of learning} \cite{transfer_learning}, whereby foundational skills in one domain (e.g., C programming) enhance students' efficiency in new contexts. Prior experience can lower {cognitive load} \cite{cognitive_load}, allowing students to focus on higher-order problem-solving rather than struggling with basic syntax. The positive relationship between PCoE and academic grades also highlights the influence of {cultural capital} \cite{cultural_capital}, wherein students from technology-rich environments begin courses with built-in advantages. 
However, these advantages expose equity concerns. The substantial performance gap between students with and without prior experience suggests a misalignment with the principles of {constructive alignment} \cite{constructive_alignment}, particularly when courses implicitly assume prior knowledge. The abrupt difficulty progression between labs and projects may also fail to respect learners' ZPD \cite{ZPD_1}, where incremental challenge and guided support are essential for skill acquisition. Lack of scaffolding disproportionately affects true beginners, leading to potential disengagement or underperformance.

The digital divide in computer and smartphone experience (PCoE, PSE) further underscores systemic inequities in access to technology \cite{digital_divide}, disproportionately impacting students from underprivileged backgrounds. These disparities are compounded by inconsistencies in TA feedback and support, undermining {assessment reliability} \cite{assessment_1}. Variations in TA training, workload, or availability may result in inequitable academic experiences across students within the same course.

The dominance of CSE (Computer Science \& Engg.) and MC (Math \& Computing) students in higher performance bands illustrates a form of {self-selection bias}, wherein students with greater prior exposure naturally gravitate toward technical disciplines. This trend creates a reinforcing feedback loop, nurturing {communities of practice} \cite{diff_theory_5} within these branches and further supporting academic success. 
In contrast, students from non-CS branches (e.g., CB, CE) tend to enter with less exposure and continue to lag in performance. The correlation between high PAE and CSE enrollment also supports the {expectancy-value theory} \cite{exp_theory}, where students are more likely to invest effort in fields they perceive as valuable and attainable based on past success.

\subsection{Comparative Study}
This subsection compares findings from the previously studied Swedish context \cite{10036401} with those from the current Indian study, focusing on engineering students’ perceptions of assessment, the roles of course coordinators and teaching assistants (TAs), and the impact of demographic and experiential factors on academic outcomes. 

In both settings, students recognized lab assignments as meaningful learning activities rather than formal requirements and viewed projects as authentic learning experiences, though they reported variations in workload across projects. Course coordinators were generally perceived as positively engaged beyond lectures, while concerns about TA consistency and availability emerged as common issues. Despite these concerns, students in both countries 
found TAs more approachable and helpful than coordinators.

A consistent trend observed across both contexts is the positive association between prior technical experience and academic performance, alongside the under-representation of female students, who, on average, performed slightly lower. However, while Swedish students reported a clearer understanding of exam expectations, Indian students' responses were more varied, indicating potential differences in communication or transparency. Swedish students also perceived a sharper difficulty increase from labs to projects.

An interesting divergence lies in support-seeking behavior: Indian students preferred consulting TAs, whereas Swedish students relied more on online resources. Additionally, the Indian study examined a broader spectrum of prior experiences, including C programming, general programming, computer, and smartphone use, and analyzed differences across academic branches, offering a more granular understanding of student backgrounds.

Overall, the results from both studies emphasize the need for greater standardization, improved TA training, and enhanced student support systems, particularly within the Indian educational context. While some challenges appear to be universal, the findings also suggest that we, as educators, can learn from each other’s strengths.

\subsection{Implications for Practice}
The findings of this study highlight several actionable strategies for course coordinators, TAs, program designers, and institutions to foster equitable, effective, and inclusive learning environments. 
First, offering {pre-course modules or bridging workshops} is vital for students with limited programming or computer experience. These initiatives should introduce foundational concepts and computational thinking to establish a common baseline before course commencement. Assignments and projects must be structured with {gradual complexity}, aligned with students’ ZPD \cite{ZPD_1}, allowing beginners to build confidence and competence over time.

To ensure {equity and consistency in assessment}, institutions should implement {regular and comprehensive TA training} \cite{TAs1}, minimizing variability in grading and feedback across different groups. Cultivating a {supportive and inclusive classroom culture} is equally important. Instructors can help mitigate stereotype threat by showcasing diverse role models, promoting a growth mindset, and fostering a sense of belonging among all students.

{Structured peer-learning activities}, such as pair programming and study groups, can enhance engagement, especially for students from underrepresented groups. Addressing {technology access gaps} is also critical. Institutions should offer {digital literacy workshops} to support students from less technology-rich backgrounds. 
Project work should be {authentic yet adaptable}, reflecting real-world problems while being accessible to students regardless of their prior experience. Offering{project options or scaffolded challenges} can help ensure meaningful participation for all.

Finally, institutions should adopt a {data-informed approach} to continuous improvement. Regular analysis of student performance and engagement, disaggregated by gender, prior experience, and academic branch, can guide targeted interventions. Collecting and acting on student feedback regarding assessment fairness, TA support, and course difficulty will further refine instructional practices. 
By implementing these strategies, institutions can help close achievement gaps, support diverse learners, and promote equitable academic success.




\subsection{Limitations}
A key limitation of this study lies in its confinement to a single Indian engineering institution, which constrains the generalizability of findings to other educational contexts with differing curricular structures, student demographics, or institutional policies. Additionally, students’ prior programming, computer, and smartphone experience were self-reported, a methodology that may introduce recall bias or inaccuracies in self-assessment of skill levels, potentially compromising the reliability of the outcomes.

Only 318 of the 516 eligible students (61.6\%) volunteered consistently in the survey, introducing potential non-response bias. Volunteers may differ systematically from non-respondents, such as being more motivated or comfortable with surveys, thus skewing the results. 

Finally, while subgroup analyses were conducted (e.g., by gender, academic branch, and prior experience), some subgroups, such as female students or students from less-represented branches, were relatively small.

\subsection{Future Research Directions} 
Future research could explore how variations in TA practices influence student learning outcomes and perceptions, particularly across different institutional or regional contexts. It would also be valuable to investigate the specific characteristics of lab assignments, exams, and projects that contribute to their perceived authenticity and effectiveness. 
Additionally, examining the experiences of subgroups such as first-generation college students or non-CS majors may reveal unique challenges or expectations. Another important direction involves studying the growing use of large language models (LLMs) by students, investigating how these tools are integrated into coursework, their impact on learning and assessment integrity, and how students and instructors perceive their value and limitations in an Indian educational context.

\section*{Acknowledgment}
The authors gratefully acknowledge the support of all students and teaching assistants in the studied course.

\begin{figure*}
\centering
\footnotesize

\begin{tabular}{c | c | c | c | c}
\hline
\includegraphics[width=0.18\linewidth]{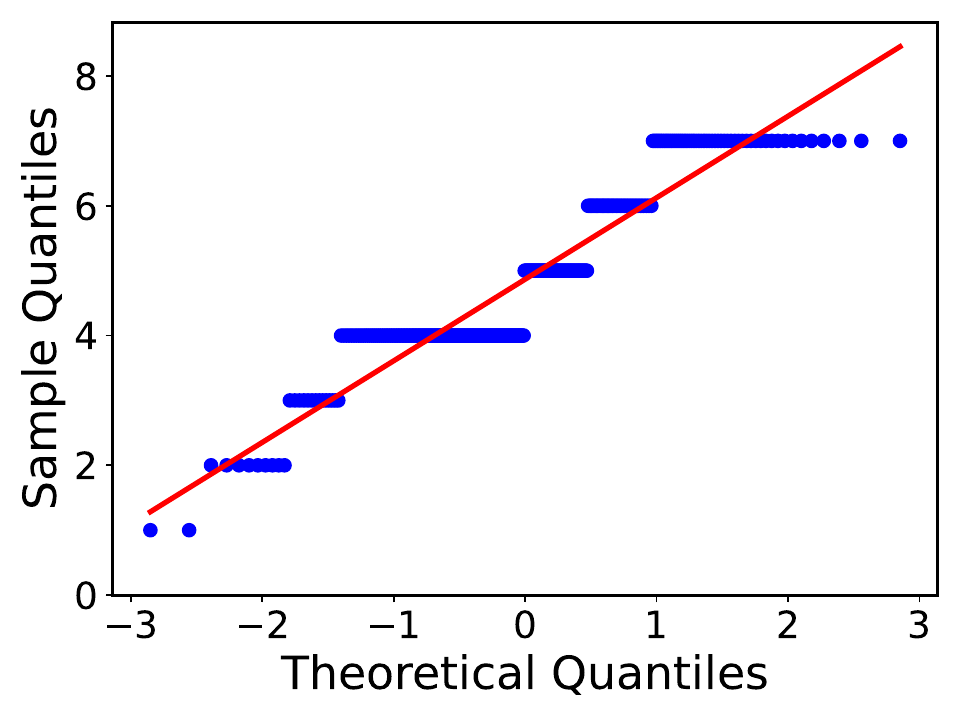} &     
\includegraphics[width=0.18\linewidth]{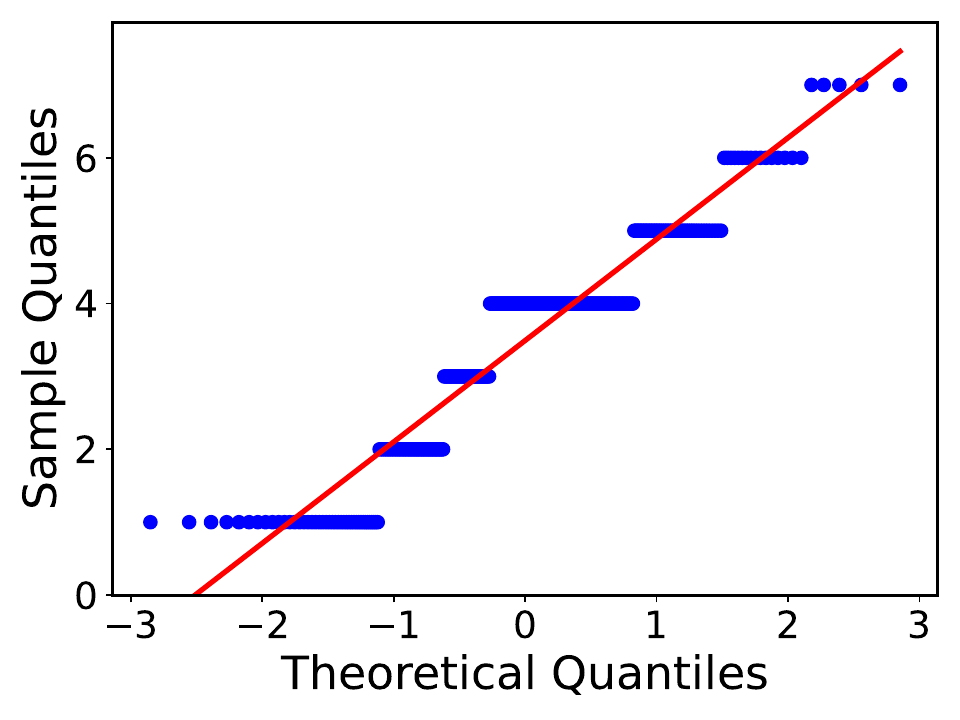} &
\includegraphics[width=0.18\linewidth]{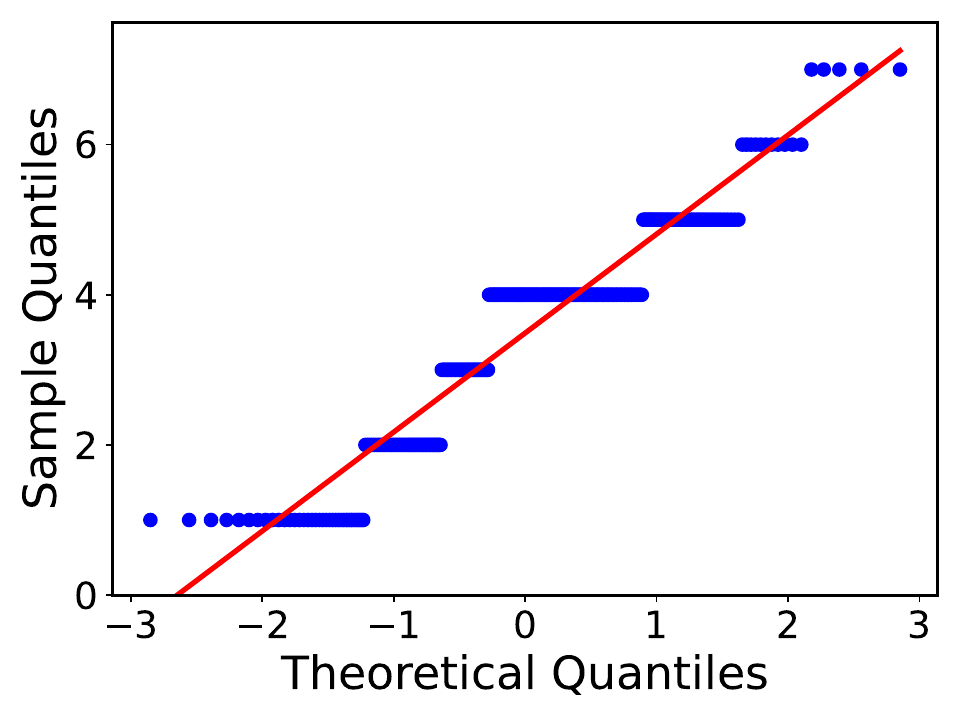} &
\includegraphics[width=0.18\linewidth]{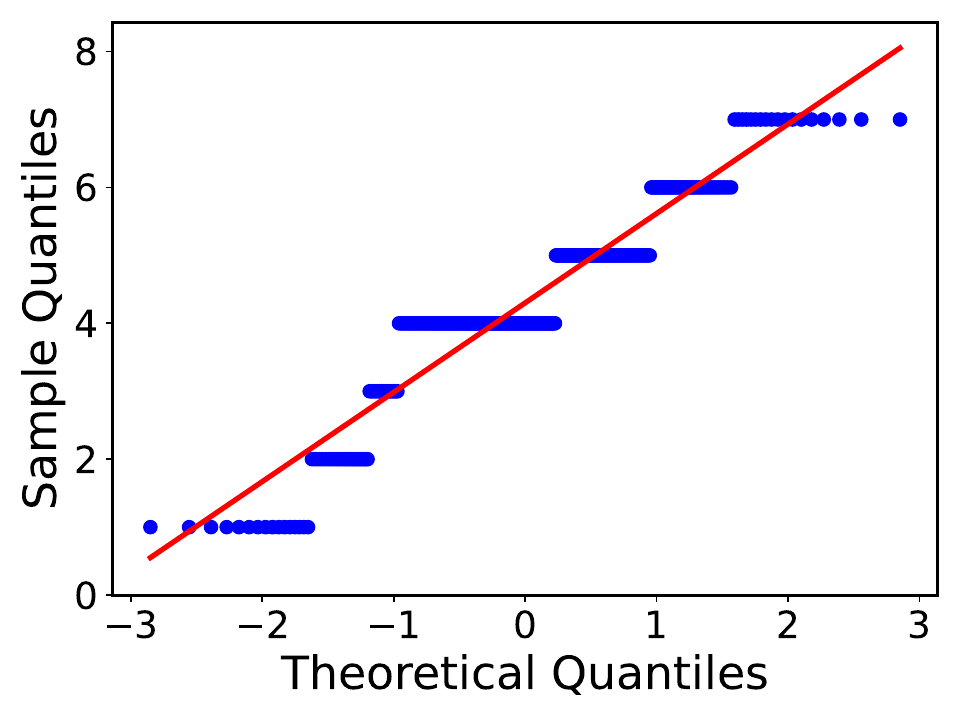} &  
\includegraphics[width=0.18\linewidth]{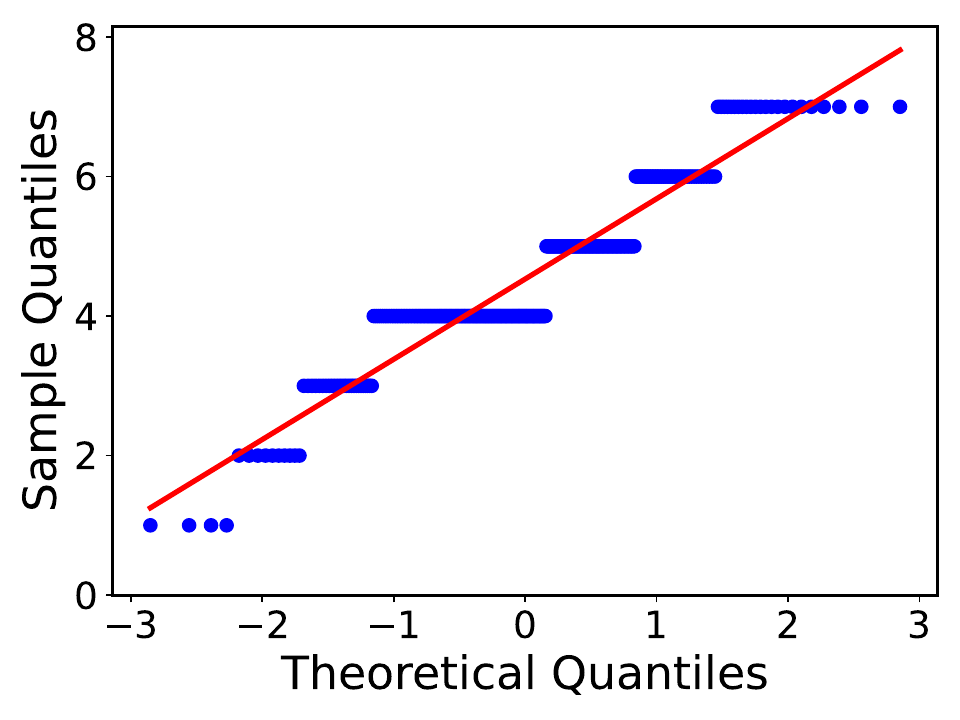} \\  
S1  & S2  & S3  & S4  & S5  \\ \hline

\includegraphics[width=0.18\linewidth]{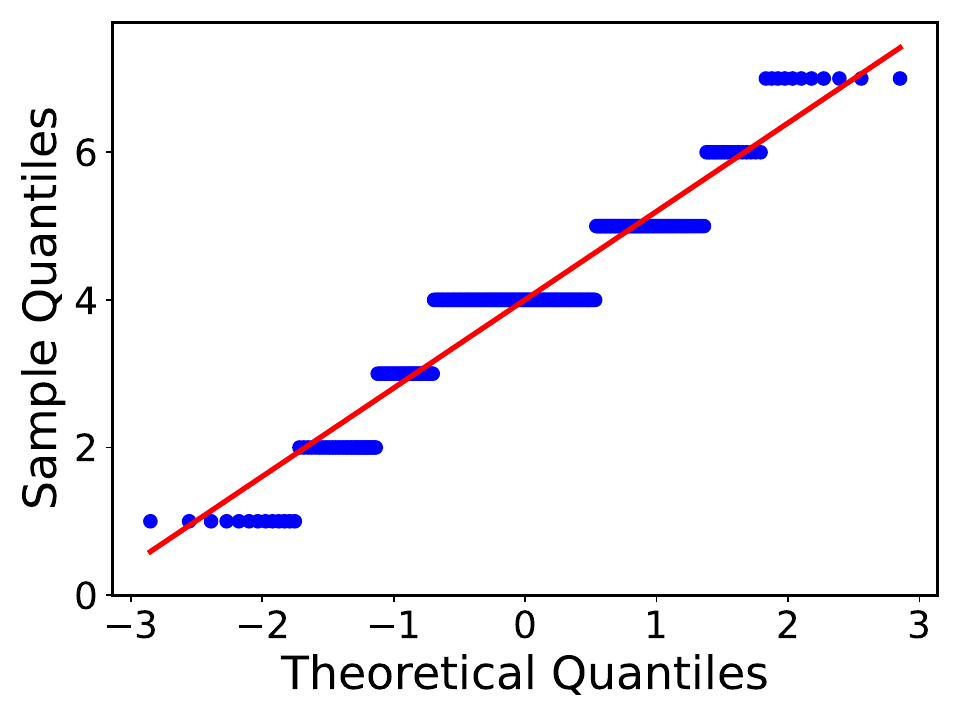} &     
\includegraphics[width=0.18\linewidth]{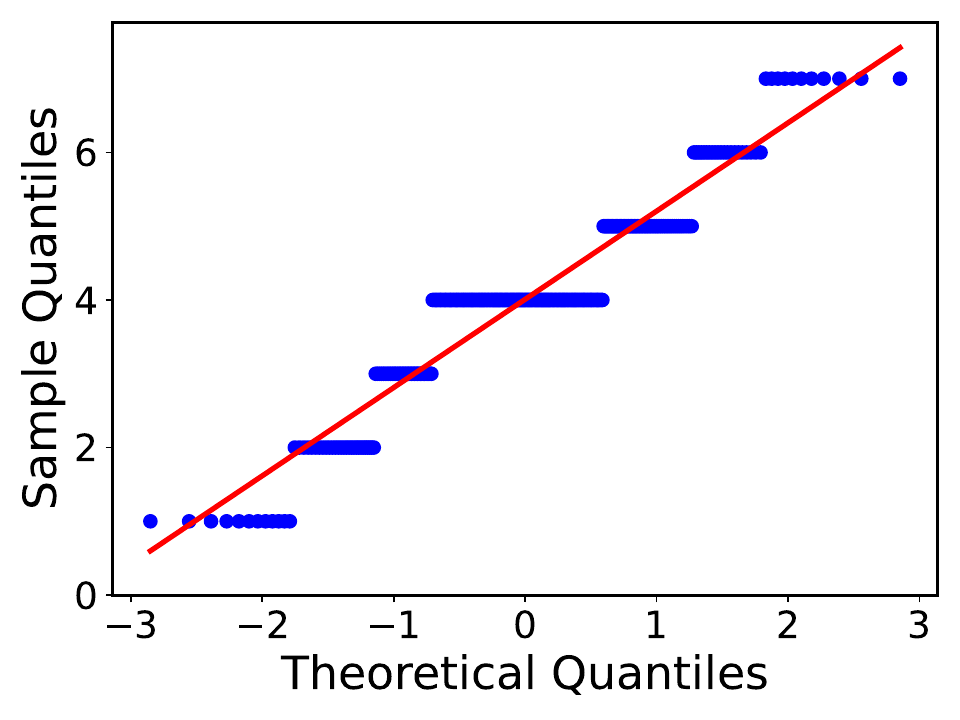} &
\includegraphics[width=0.18\linewidth]{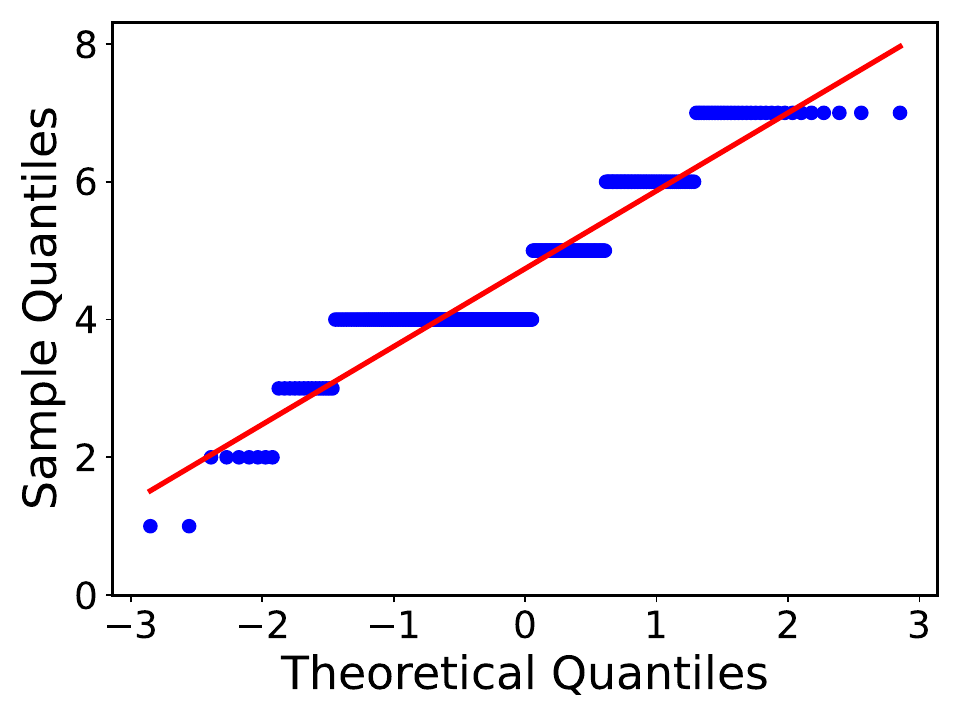} &
\includegraphics[width=0.18\linewidth]{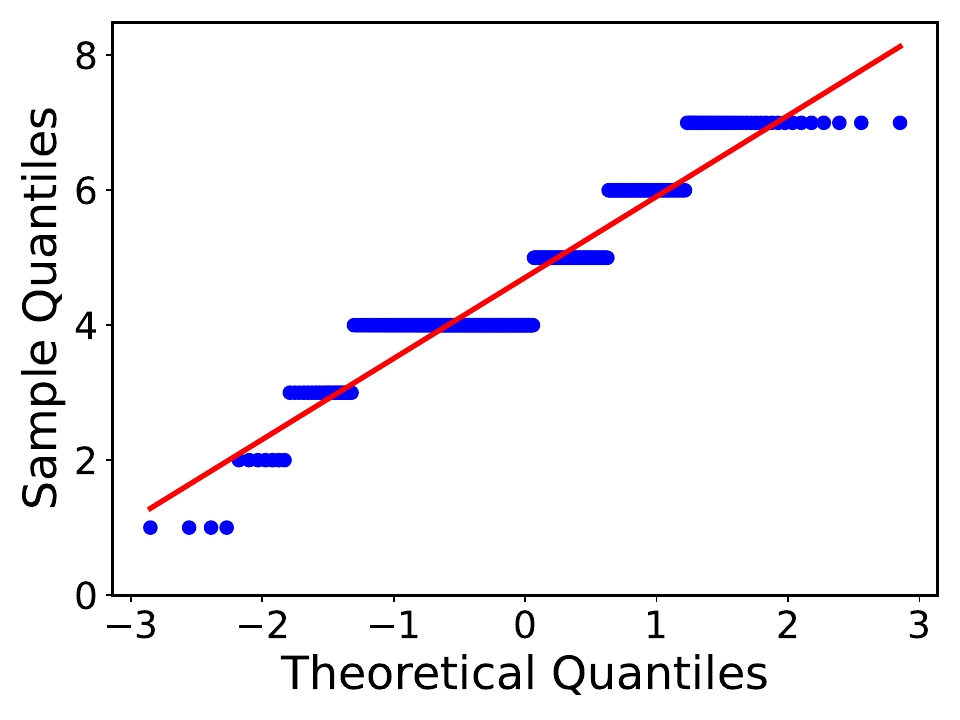} &  
\includegraphics[width=0.18\linewidth]{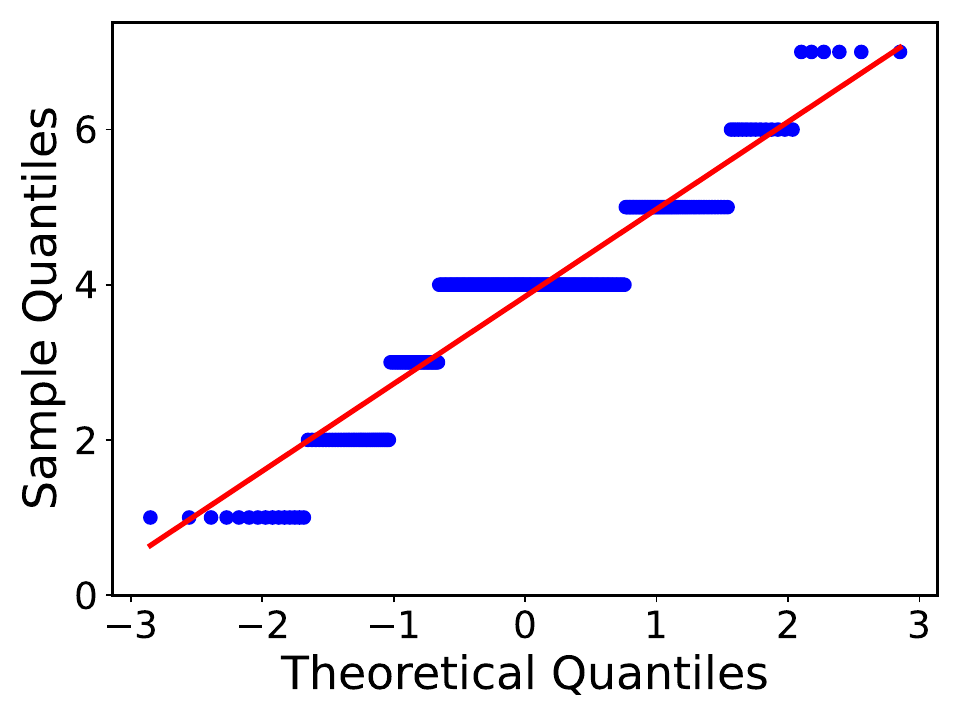} \\  
S6  & S7  & S8  & S9  & S10 \\ \hline

\includegraphics[width=0.18\linewidth]{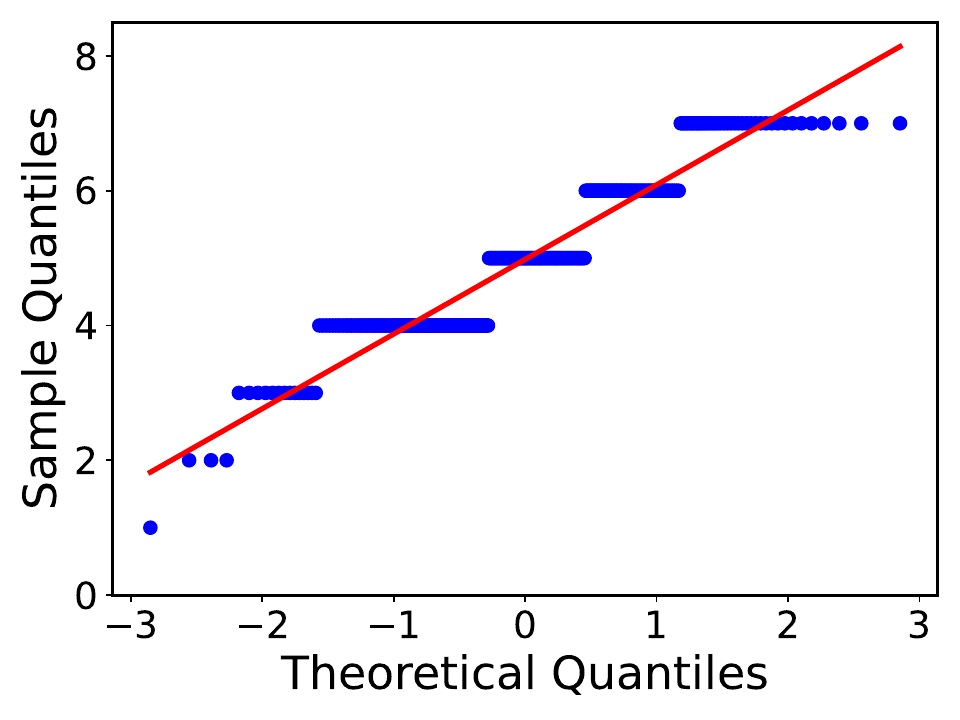} &     
\includegraphics[width=0.18\linewidth]{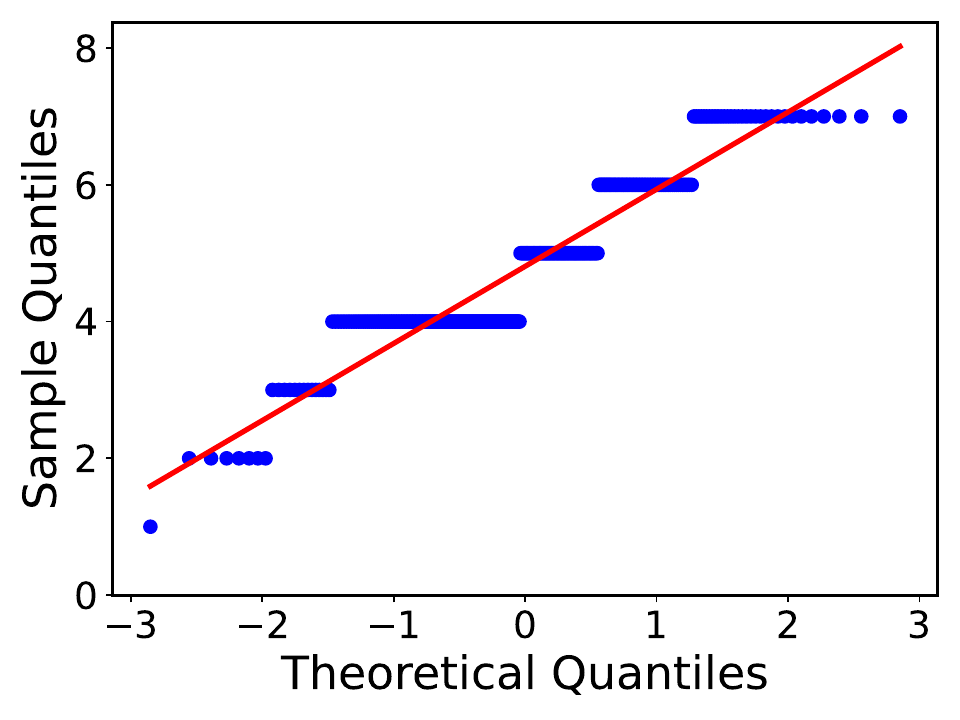} &
\includegraphics[width=0.18\linewidth]{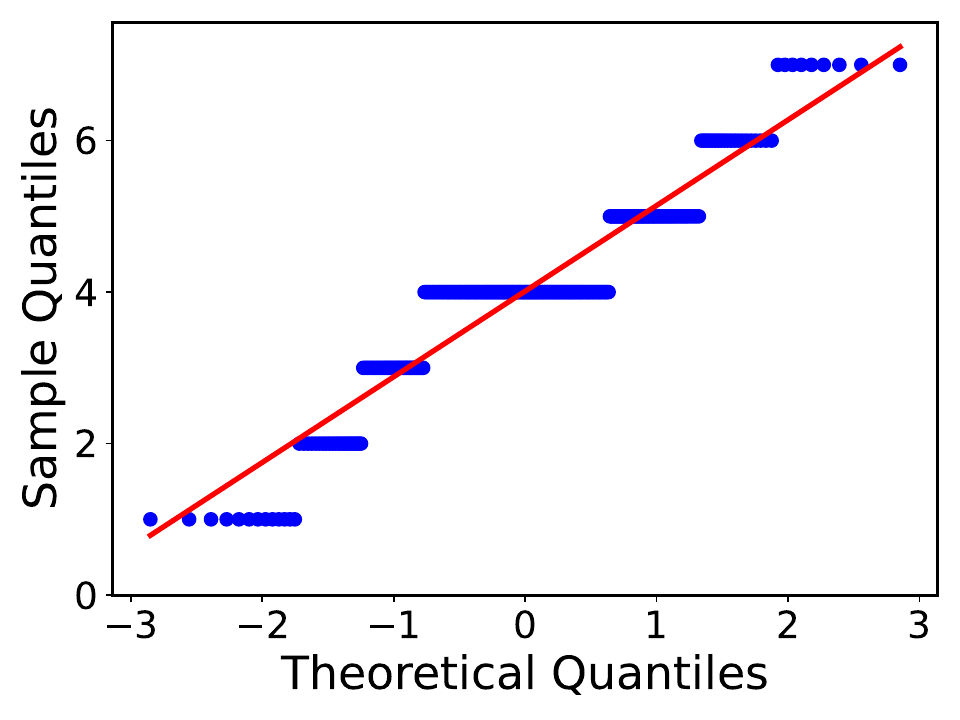} &
\includegraphics[width=0.18\linewidth]{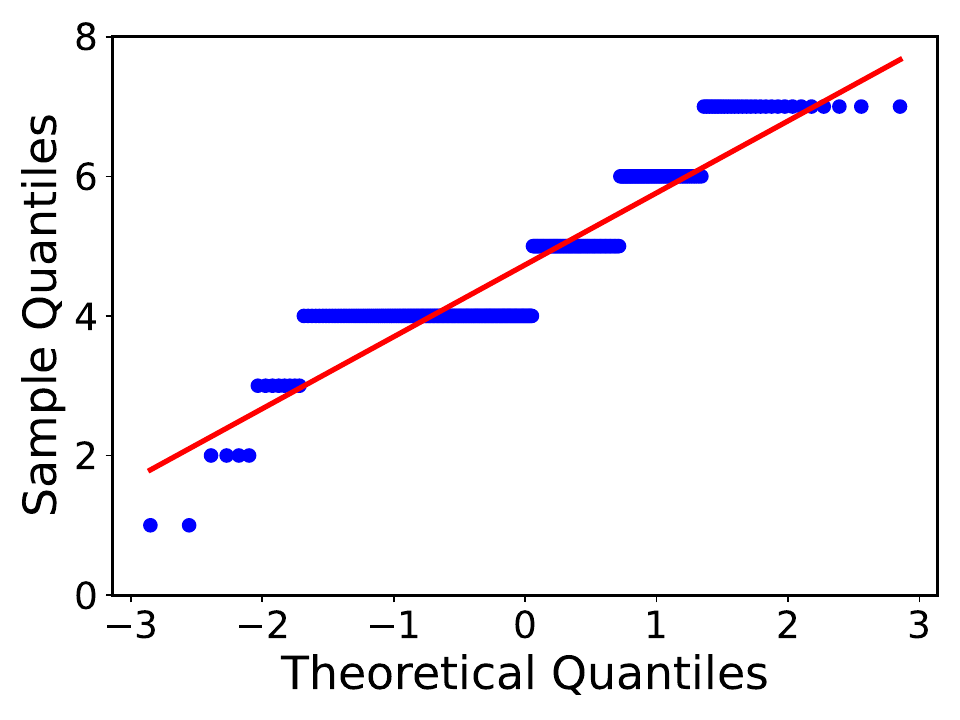} &  
\includegraphics[width=0.18\linewidth]{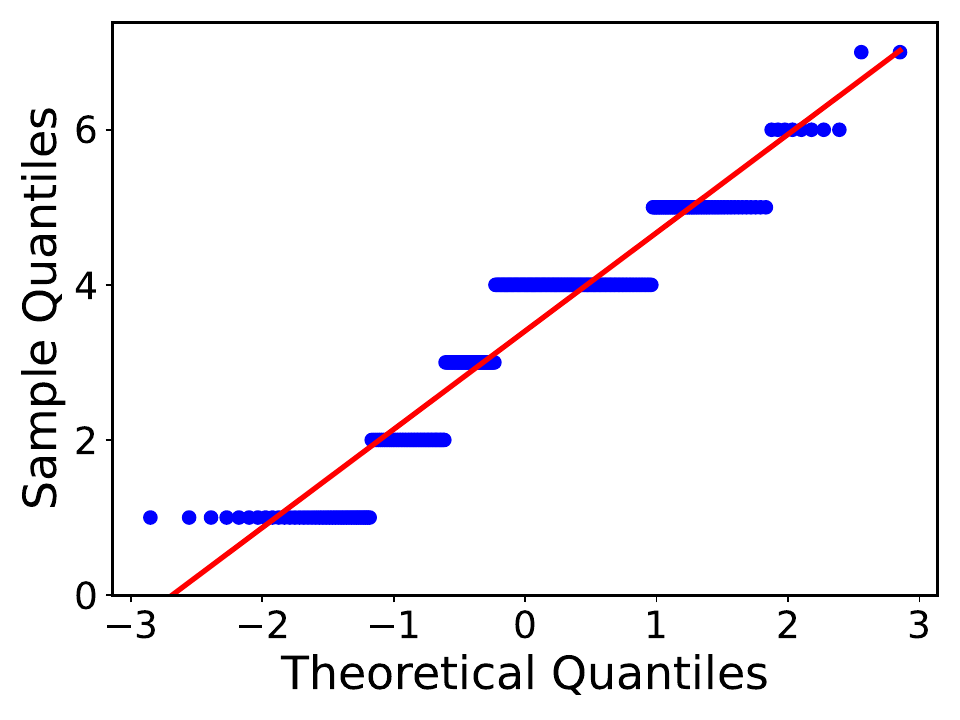} \\  
S11 & S12 & S13 & S14 & S15 \\ \hline

\includegraphics[width=0.18\linewidth]{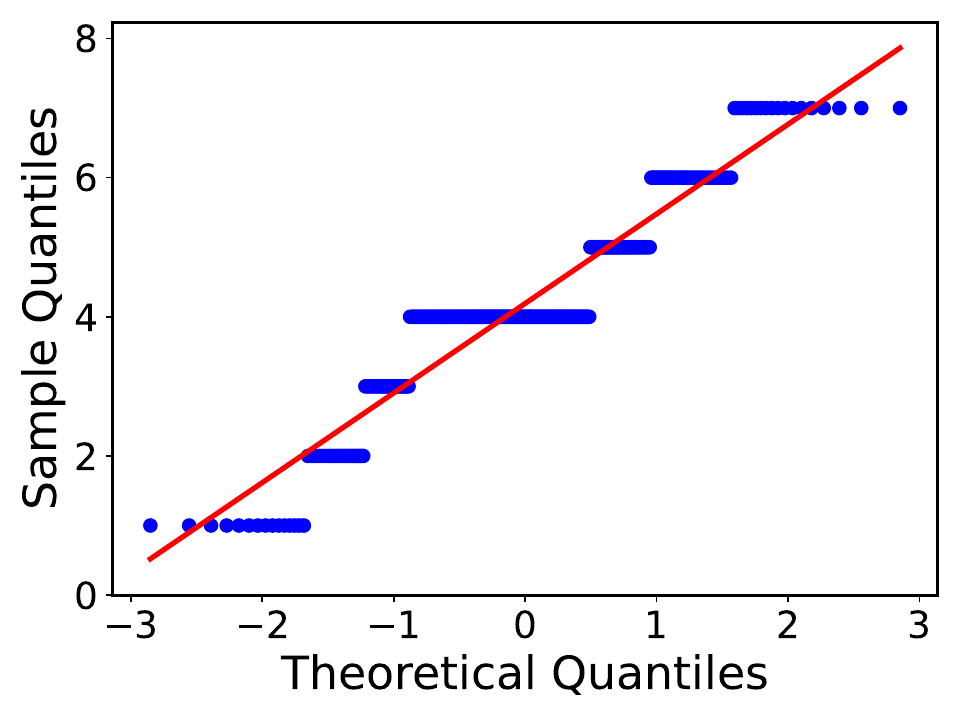} &     
\includegraphics[width=0.18\linewidth]{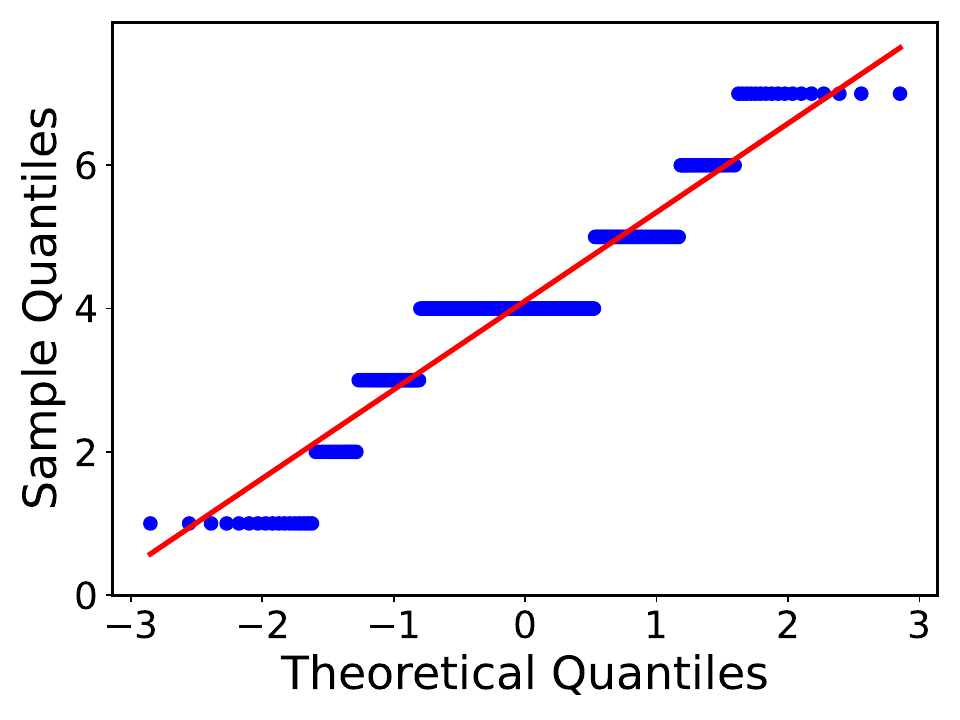} &
\includegraphics[width=0.18\linewidth]{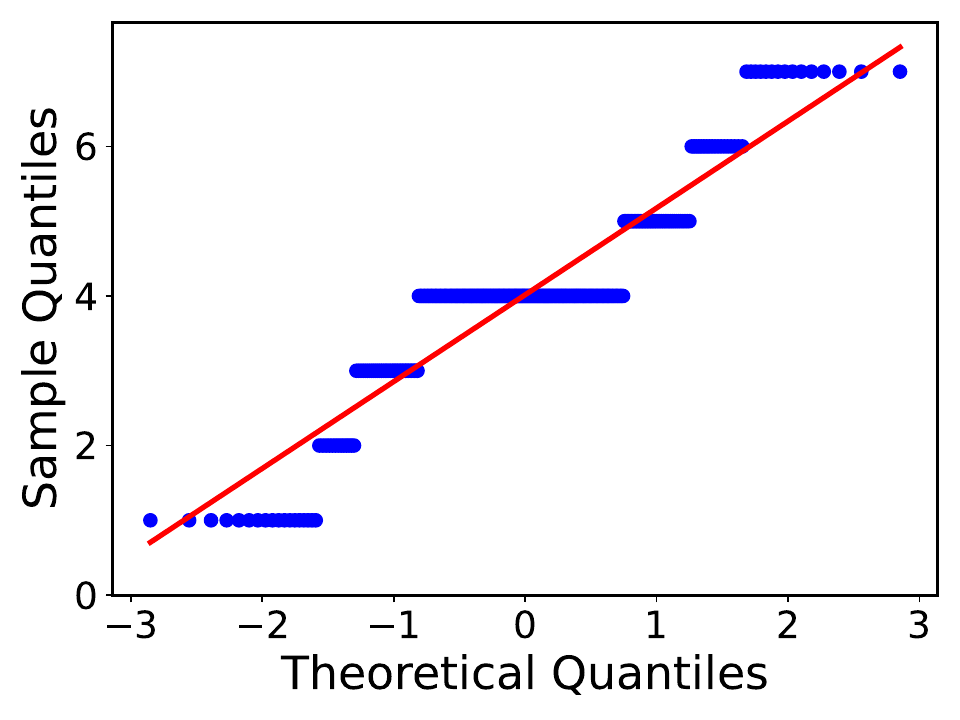} &
\includegraphics[width=0.18\linewidth]{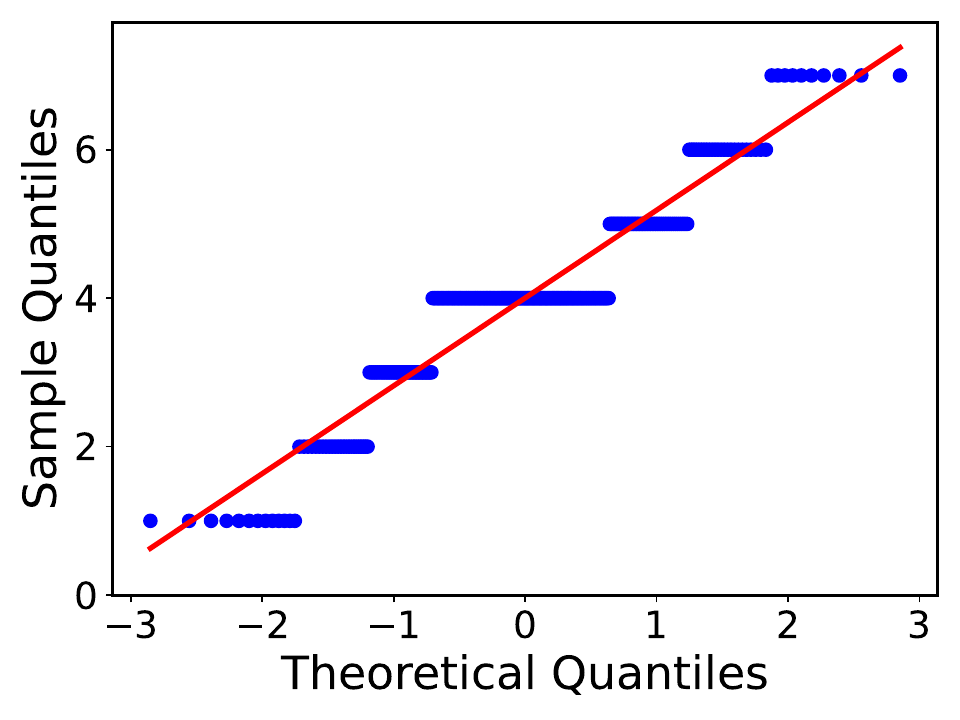} &  
\includegraphics[width=0.18\linewidth]{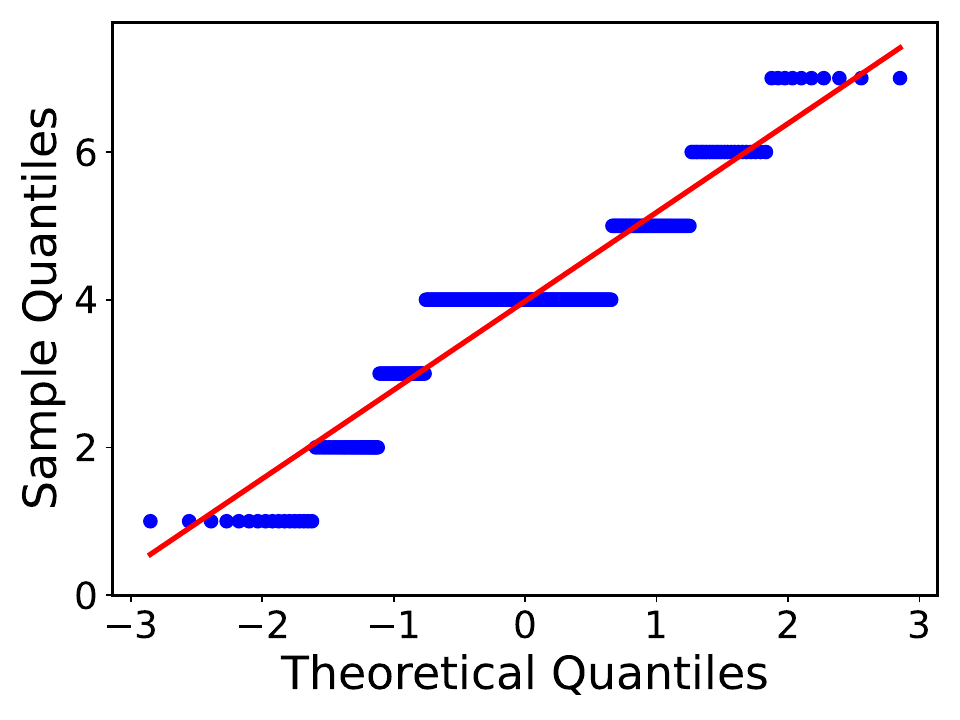} \\  
S16 & S17 & S18 & S19 & S20 \\ \hline

\includegraphics[width=0.18\linewidth]{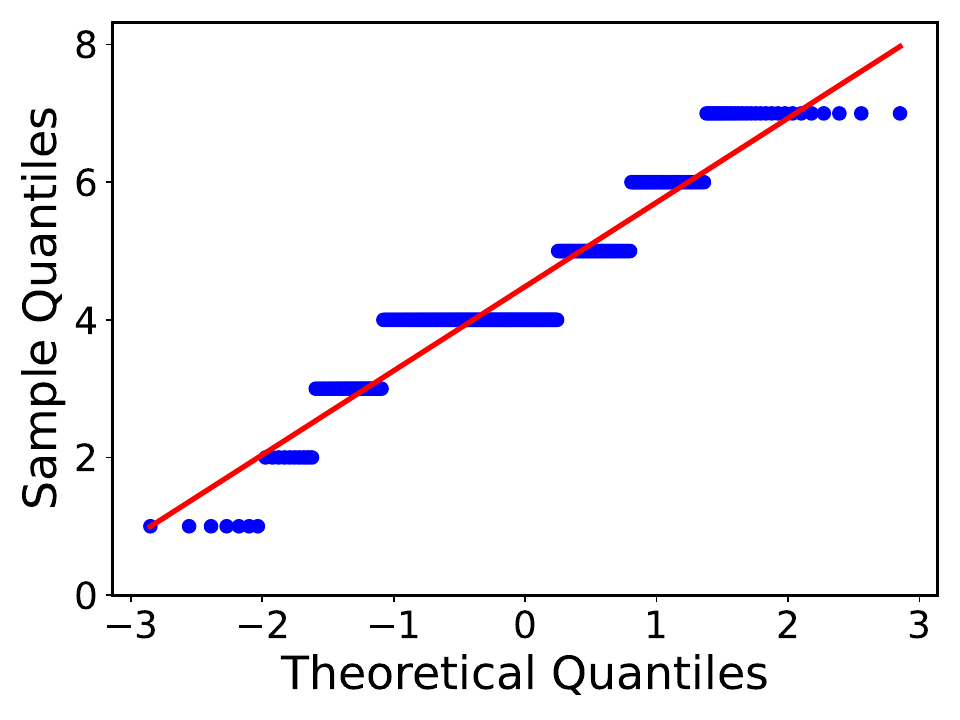} &     
\includegraphics[width=0.18\linewidth]{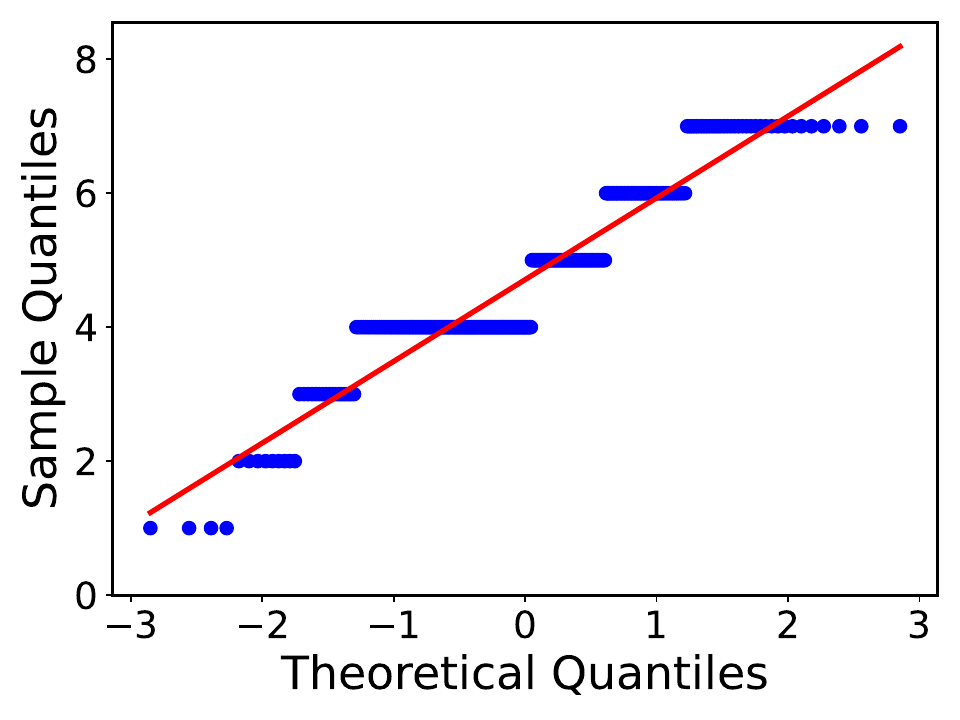} &
\includegraphics[width=0.18\linewidth]{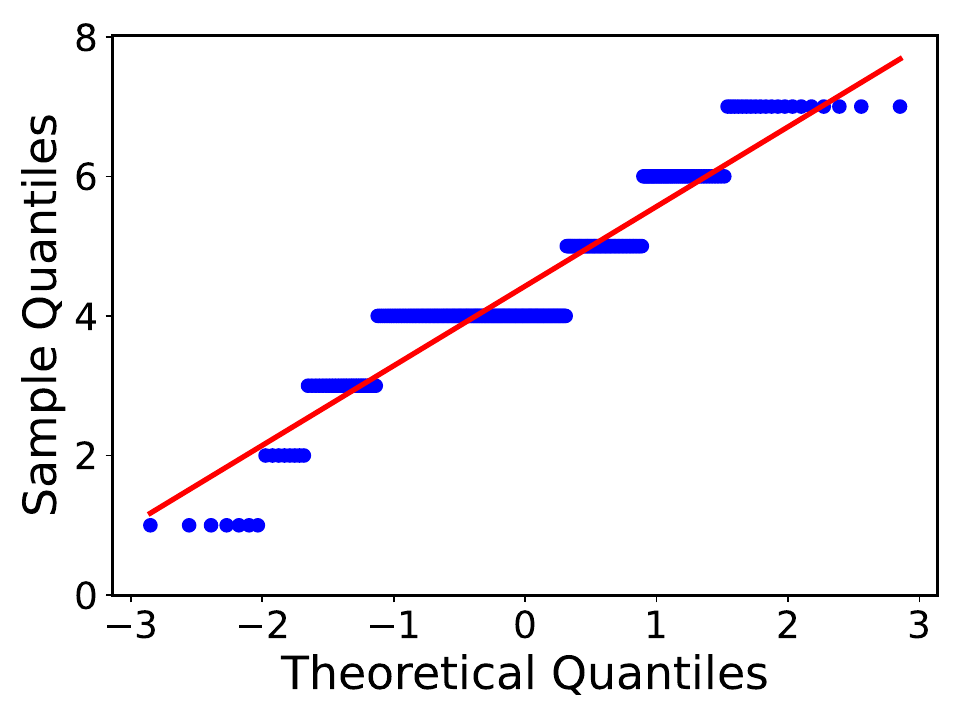} &
\includegraphics[width=0.18\linewidth]{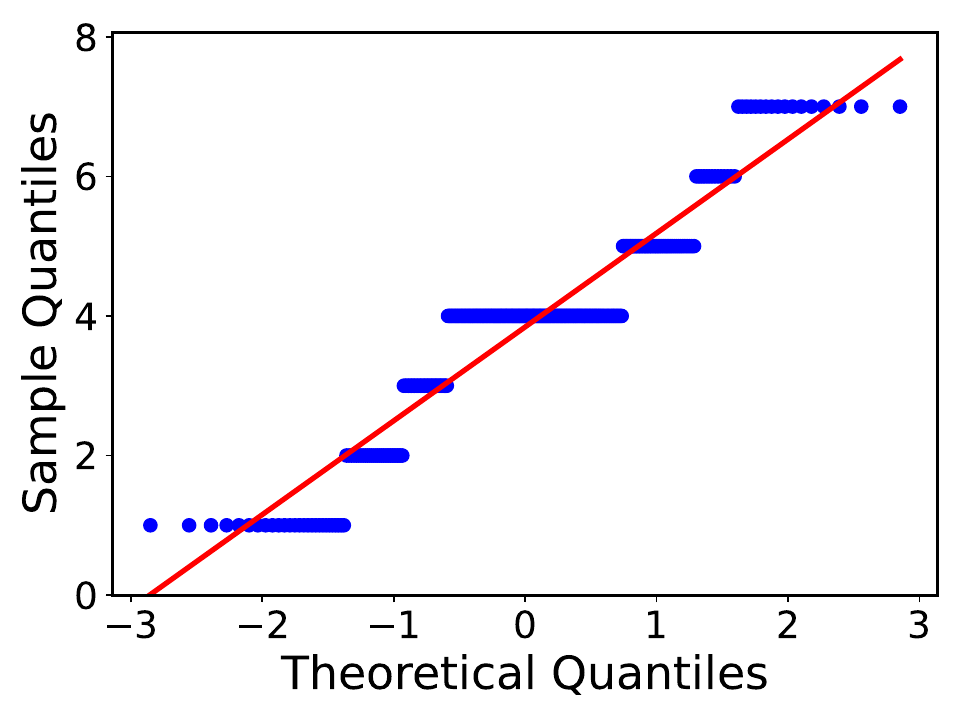} &  
\includegraphics[width=0.18\linewidth]{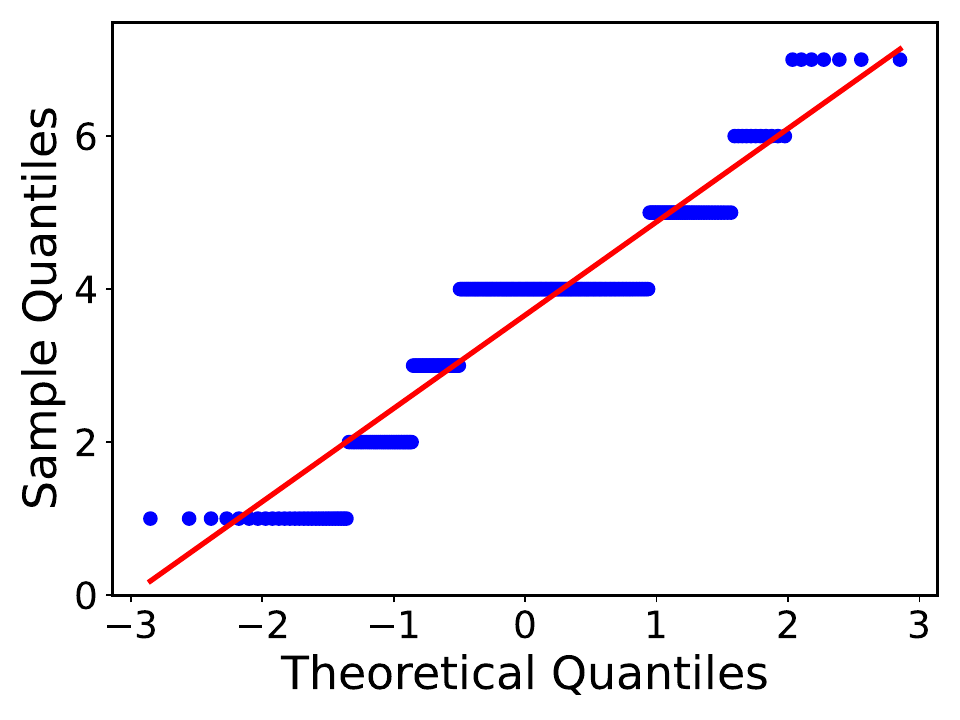} \\  
S21 & S22 & S23 & S24 & S25 \\
\hline
\end{tabular}

\caption{Q-Q plots for survey questions S1 to S25 from Week-9. The plots indicate that response distributions deviate from normality, thereby justifying the application of non-parametric statistical tests in the analysis.}

\label{fig:QQ}
\end{figure*}

\balance 
\bibliographystyle{IEEEtran}  
\bibliography{ref.bib}

\clearpage
\appendices




\section{Q-Q Plots for Week-9} \label{App:qq_plots}
Analysis of the Q-Q plots presented in Fig. \ref{fig:QQ} for student responses to all 25 Week-9 statements indicates that the data for each question deviates from a normal distribution. Due to this non-normality, non-parametric statistical methods, specifically the Kruskal-Wallis and Mann-Whitney U tests, were employed for the Week-9 analysis, as they are well-suited for such data.

\section{Analyses of Initial Weeks} \label{App:weeks_analysis}
For the initial eight weeks, Tables~\ref{tab:week0_analysis}--\ref{tab:week11_analysis} present the week-wise $p$-value analysis using the Mann-Whitney U test for comparing subgroups based on gender (G), and the Kruskal-Wallis test for comparing subgroups based on 
academic performance grade (Gd), branch (B), and
prior experiences in C programming (PCE)/ 
any programming language (PAE)/ 
computer usage (PCoE)/ 
smartphone usage (PSE). 
Here, $p$-values indicating strong significance ($p \leq 0.05$) are shown in \textcolor{darkred}{\textbf{red}} color.

\begin{table}[!b]
\centering
\caption{Analysis of $p$-values for Week-1}
\begin{adjustbox}{width=0.45\textwidth} 
\begin{tabular}{ll|c|c|c|c|c|c|c}
\hline
\textbf{SN.} & \textbf{Statement} & \textbf{G} & \textbf{PCE} & \textbf{PAE} & \textbf{PCoE} & \textbf{PSE} & \textbf{Gd} & \textbf{B} \\ \hline

S1: & Learning activity & 0.491 & \textcolor{darkred}{\textbf{0.022}} & 0.335 & 0.170 & 0.751 & 0.947 & 0.974 \\ 
S2: & Necessary evil & 0.543 & 0.860 & 0.838 & 0.749 & 0.571 & 0.930 & 0.322 \\ 
S3: & Stressful & 0.829 & \textcolor{darkred}{\textbf{0.036}} & 0.244 & \textcolor{darkred}{\textbf{0.016}} & 0.988 & 0.299 & 0.693 \\ 
S4: & Help from peers & 0.458 & 0.690 & \textcolor{darkred}{\textbf{0.029}} & 0.279 & 0.221 & \textcolor{darkred}{\textbf{0.004}} & 0.342 \\ 
S5: & Help to peers & 0.342 & 0.119 & 0.597 & \textcolor{darkred}{\textbf{0.001}} & 0.267 & 0.112 & 0.475 \\ \hline

S6: & Predictable question types & 0.432 & 0.244 & \textcolor{darkred}{\textbf{0.017}} & \textcolor{darkred}{\textbf{0.007}} & 0.565 & 0.339 & 0.134 \\ 
S7: & Predictable assessment & 0.110 & 0.288 & 0.093 & 0.056 & 0.135 & 0.529 & 0.058 \\ 
S8: & Skills and knowledge & 0.378 & 0.135 & 0.232 & 0.388 & 0.340 & 0.704 & 0.783 \\ 
S9: & Authentic & 0.893 & 0.179 & \textcolor{darkred}{\textbf{0.038}} & 0.439 & 0.362 & 0.305 & 0.581 \\ \hline

S10: & Big leap & 0.951 & 0.208 & 0.370 & 0.600 & 0.821 & 0.079 & \textcolor{darkred}{\textbf{0.047}} \\ 
S11: & Learnt a lot & 0.742 & 0.135 & 0.173 & 0.211 & 0.963 & 0.864 & 0.221 \\ 
S12: & Authentic & 0.886 & 0.093 & \textcolor{darkred}{\textbf{0.018}} & 0.112 & 0.465 & 0.372 & 0.444 \\ 
S13: & Difference & 0.589 & 0.123 & \textcolor{darkred}{\textbf{0.023}} & 0.205 & 0.130 & 0.464 & 0.131 \\ \hline

S14: & Course structure & 0.300 & 0.057 & \textcolor{darkred}{\textbf{0.013}} & 0.427 & 0.251 & 0.329 & 0.394 \\ 
S15: & Little insight & 0.370 & 0.065 & 0.756 & 0.052 & 0.240 & 0.520 & 0.602 \\ \hline

S16: & Assessment & 0.507 & 0.728 
& 0.179 & 0.142 & 0.223 & 0.560 & 0.111 \\ 
S17: & Variation in lab assessment & 0.597 & 0.353 & 0.468 & \textcolor{darkred}{\textbf{0.021}} & 0.668 & 0.909 & 0.229 \\ 
S18: & Variation for projects & 0.803 & 0.774 & 0.503 & \textcolor{darkred}{\textbf{0.012}} & 0.737 & 0.735 & 0.417 \\ 
S19: & Variation for help & \textcolor{darkred}{\textbf{0.012}} & 0.113 & 0.554 & \textcolor{darkred}{\textbf{0.004}} & 0.809 & 0.952 & 0.314 \\ 
S20: & Tutorials & 0.232 & 0.500 & 0.678 & 0.065 & 0.600 & 0.106 & 0.113 \\ 
S21: & Professionalism & 0.571 & 0.737 & 0.124 & 0.062 & \textcolor{darkred}{\textbf{0.024}} & 0.711 & 0.589 \\
S22: & Feedback & 0.768 & 0.561 & 0.261 & 0.376 & 0.683 & 0.801 & 0.171 \\
S23: & Asking TA help & 0.098 & 0.118 & 0.220 & 0.095 & 0.472 & 0.823 & \textcolor{darkred}{\textbf{0.008}} \\
S24: & Preferred internet & \textcolor{darkred}{\textbf{0.036}} & 0.688 & \textcolor{darkred}{\textbf{0.043}} & 0.183 & 0.281 & 0.152 & 0.563 \\ 
S25: & Few TAs & \textcolor{darkred}{\textbf{0.023}} & 0.307 & 0.587 & \textcolor{darkred}{\textbf{0.004}} & 0.520 & 0.696 & 0.519 \\ \hline
\end{tabular}
\end{adjustbox}
\label{tab:week0_analysis}
\end{table}

\begin{table}[!b]
\centering
\caption{Analysis of $p$-values for Week-2}
\begin{adjustbox}{width=0.45\textwidth} 
\begin{tabular}{ll|c|c|c|c|c|c|c}
\hline
\textbf{SN.} & \textbf{Statement} & \textbf{G} & \textbf{PCE} & \textbf{PAE} & \textbf{PCoE} & \textbf{PSE} & \textbf{Gd} & \textbf{B} \\ \hline
S1: & Learning activity & 0.084 & 0.136 & \textcolor{darkred}{\textbf{0.050}} & 0.570 & 0.184 & 0.858 & 0.494 \\ 
S2: & Necessary evil & 0.871 & \textcolor{darkred}{\textbf{0.023}} & 0.654 & 0.536 & 0.866 & 0.659 & 0.399 \\ 
S3: & Stressful & 0.715 & 0.239 & \textcolor{darkred}{\textbf{0.070}} & 0.853 & 0.935 & 0.339 & \textcolor{darkred}{\textbf{0.022}} \\ 
S4: & Help from peers & 0.279 & 0.336 & 0.066 & 0.165 & \textcolor{darkred}{\textbf{0.042}} & 0.070 & 0.734 \\ 
S5: & Help to peers & 0.505 & 0.486 & 0.331 & \textcolor{darkred}{\textbf{0.006}} & 0.474 & 0.762 & 0.171 \\ \hline

S6: & Predictable question types & 0.279 & 0.466 & \textcolor{darkred}{\textbf{0.001}} & \textcolor{darkred}{\textbf{0.009}} & 0.080 & 0.507 & \textcolor{darkred}{\textbf{0.022}} \\ 
S7: & Predictable assessment & 0.453 & 0.492 & \textcolor{darkred}{\textbf{0.009}} & 0.150 & 0.184 & 0.481 & 0.087 \\ 
S8: & Skills and knowledge & 0.105 & 0.766 & 0.548 & 0.782 & 0.557 & 0.774 & 0.986 \\ 
S9: & Authentic & 0.183 & 0.090 & \textcolor{darkred}{\textbf{0.006}} & 0.623 & \textcolor{darkred}{\textbf{0.026}} & 0.301 & 0.600 \\ \hline

S10: & Big leap & 0.913 & 0.058 & 0.829 & 0.939 & 0.192 & 0.262 & \textcolor{darkred}{\textbf{0.043}} \\ 
S11: & Learnt a lot & \textcolor{darkred}{\textbf{0.030}} & 0.644 & 0.524 & 0.931 & 0.684 & 0.480 & 0.149 \\ 
S12: & Authentic & 0.807 & 0.398 & 0.427 & 0.636 & 0.894 & 0.195 & 0.467 \\ 
S13: & Difference & 0.738 & 0.415 & 0.447 & 0.201 & 0.590 & 0.065 & 0.077 \\ \hline

S14: & Course structure & 0.508 & \textcolor{darkred}{\textbf{0.040}} & 0.744 & 0.123 & 0.468 & 0.367 & 0.825 \\ 
S15: & Little insight & 0.955 & 0.338 & 0.060 & 0.323 & 0.715 & \textcolor{darkred}{\textbf{0.003}} & \textcolor{darkred}{\textbf{0.038}} \\ \hline

S16: & Assessment & 0.633 & 0.770 & 0.944 & 0.904 & 0.239 & 0.716 & \textcolor{darkred}{\textbf{0.039}} \\ 
S17: & Variation in lab assessment & 0.263 & 0.417 & 0.871 & 0.465 & 0.519 & 0.871 & 0.141 \\ 
S18: & Variation for project & 0.272 & 0.143 & 0.569 & 0.260 & 0.171 & 0.636 & 0.092 \\ 
S19: & Variation for help & 0.873 & 0.142 & 0.715 & 0.367 & 0.822 & 0.532 & 0.139 \\ 
S20: & Tutorials & 0.624 & 0.224 & 0.115 & 0.880 & 0.827 & 0.558 & 0.110 \\
S21: & Professionalism & 0.755 & 0.516 & 0.428 & 0.062 & \textcolor{darkred}{\textbf{0.003}} & 0.551 & 0.142 \\
S22: & Feedback & 0.745 & 0.400 & 0.691 & 0.364 & 0.084 & 0.846 & 0.676 \\ 
S23: & Asking TA help & 0.151 & 0.470 & 0.776 & 0.425 &0.055 & 0.854 & 0.979 \\ 
S24: & Preferred internet& 0.987 & 0.512 & 0.051 & \textcolor{darkred}{\textbf{0.046}} & 0.124 & 0.421 & 0.350 \\ 
S25: & Few TAs & 0.564 & 0.506 & 0.885 & 0.216 & 0.897 & \textcolor{darkred}{\textbf{0.026}} & 0.087 \\ \hline

\end{tabular}
\label{tab:week3_analysis}
\end{adjustbox}
\end{table}

\begin{table}[]
\centering
\caption{Analysis of $p$-values for Week-3}
\begin{adjustbox}{width=0.45\textwidth} 
\begin{tabular}{ll|c|c|c|c|c|c|c}
\hline
\textbf{SN.} & \textbf{Statement} & \textbf{G} & \textbf{PCE} & \textbf{PAE} & \textbf{PCoE} & \textbf{PSE} & \textbf{Gd} & \textbf{B} \\
\hline
S1: & Learning activity & 0.591 & \textcolor{darkred}{\textbf{0.033}} & \textcolor{darkred}{\textbf{0.003}} & 0.131 & 0.579 & 0.261 & 0.615 \\
S2: & Necessary evil & 0.197 & 0.069 & 0.599 & \textcolor{darkred}{\textbf{0.010}} & 0.177 & 0.490 & \textcolor{darkred}{\textbf{0.028}} \\
S3: & Stressful & 0.449 & 0.092 & \textcolor{darkred}{\textbf{<0.001}} & 0.057 & 0.162 & 0.159 & 0.497 \\
S4: & Help from peers & 0.570 & 0.246 & \textcolor{darkred}{\textbf{0.010}} & 0.195 & 0.249 & 0.128 & \textcolor{darkred}{\textbf{0.050}} \\
S5: & Help to peers & 0.886 & 0.658 & 0.144 & 0.406 & 0.372 & \textcolor{darkred}{\textbf{0.009}} & 0.943 \\
\hline

S6: & Predictable question types & 0.618 & 0.214 & 0.280 & 0.513 & 0.178 & 0.840 & 0.114 \\
S7: & Predictable assessment & 0.547 & 0.316 & 0.433 & 0.326 & 0.094 & 0.309 & 0.064 \\
S8: & Skills and knowledge & 0.487 & 0.172 & \textcolor{darkred}{\textbf{0.010}} & 0.895 & 0.437 & 0.780 & 0.571 \\
S9: & Authentic & 0.812 & 0.232 & \textcolor{darkred}{\textbf{<0.001}} & 0.241 & 0.214 & 0.455 & 0.183 \\
\hline

S10: & Big leap & 0.550 & \textcolor{darkred}{\textbf{0.026}} & 0.483 & 0.197 & 0.326 & 0.769 & 0.286 \\
S11: & Learnt a lot & 0.420 & \textcolor{darkred}{\textbf{0.013}} & \textcolor{darkred}{\textbf{0.020}} & \textcolor{darkred}{\textbf{0.048}} & 0.373 & 0.696 & 0.453 \\
S12: & Authentic & 0.835 & \textcolor{darkred}{\textbf{0.037}} & \textcolor{darkred}{\textbf{0.028}} & \textcolor{darkred}{\textbf{0.048}} & 0.307 & 0.832 & 0.076 \\
S13: & Difference & 0.852 & 0.080 & 0.340 & 0.757 & 0.623 & 0.149 & 0.077 \\
\hline

S14: & Course structure & 0.913 & 0.308 & \textcolor{darkred}{\textbf{<0.001}} & \textcolor{darkred}{\textbf{0.009}} & 0.260 & 0.179 & 0.629 \\
S15: & Little insight & 0.483 & 0.374 & \textcolor{darkred}{\textbf{0.004}} & \textcolor{darkred}{\textbf{0.032}} & 0.380 & 0.166 & \textcolor{darkred}{\textbf{0.044}} \\
\hline

S16: & Assessment & 0.115 & 0.108 & 0.071 & 0.274 & 0.107 & 0.295 & 0.651 \\
S17: & Variation in lab assessment & 0.675 & 0.436 & 0.769 & 0.949 & 0.677 & 0.416 & 0.292 \\
S18: & Variation for project & 0.746 & 0.325 & 1.000 & 0.666 & 0.781 & 0.546 & 0.084 \\
S19: & Variation for help & 0.674 & 0.297 & 0.608 & 0.699 & 0.726 & 0.813 & 0.897 \\
S20: & Tutorials & 0.746 & 0.941 & 0.789 & \textcolor{darkred}{\textbf{0.027}} & 0.696 & 0.257 & 0.802 \\
S21: & Professionalism & 0.946 & 0.676 & 0.226 & 0.209 & \textcolor{darkred}{\textbf{0.013}} & 0.194 & 0.230 \\
S22: & Feedback & 0.595 & 0.185 & 0.055 & 0.100 & 0.412 & 0.321 & 0.706 \\
S23: & Asking TA help & 0.251 & 0.852 & 0.360 & 0.737 & 0.743 & 0.454 & 0.412 \\
S24: & Preferred internet & 0.311 & 0.964 & 0.213 & 0.226 & 0.125 & 0.760 & 0.197 \\
S25: & Few TAs & 0.516 & 0.570 & 0.149 & \textcolor{darkred}{\textbf{0.020}} & 0.689 & 0.292 & 0.189 \\
\hline

\end{tabular}
\label{tab:week6_analysis}
\end{adjustbox}
\end{table}

\begin{table}[]
\centering
\caption{Analysis of $p$-values for Week-4}
\begin{adjustbox}{width=0.45\textwidth} 
\begin{tabular}{ll|c|c|c|c|c|c|c}
\hline
\textbf{SN.} & \textbf{Statement} & \textbf{G} & \textbf{PCE} & \textbf{PAE} & \textbf{PCoE} & \textbf{PSE} & \textbf{Gd} & \textbf{B} \\
\hline
S1: & Learning activity & 0.424 & 0.285 & \textcolor{darkred}{\textbf{0.031}} & 0.171 & 0.491 & 0.734 & 0.644 \\

S2: & Necessary evil & 0.987 & 0.117 & 0.052 & 0.559 & 0.710 & 0.520 & 0.228 \\

S3: & Stressful & 0.440 & 0.141 & \textcolor{darkred}{\textbf{<0.001}} & 0.883 & 0.144 & 0.844 & 0.515 \\

S4: & Help from peers & \textcolor{darkred}{\textbf{0.005}} & 0.270 & \textcolor{darkred}{\textbf{0.005}} & 0.374 & 0.280 & 0.167 & 0.630 \\

S5: & Help to peers & 0.673 & 0.718 & \textcolor{darkred}{\textbf{0.047}} & 0.060 & 0.135 & 0.271 & 0.710 \\
\hline

S6: & Predictable question types & 0.245 & 0.571 & 0.751 & 0.541 & 0.131 & 0.764 & 0.542 \\
S7: & Predictable assessment & 0.098 & 0.331 & 0.312 & 0.639 & 0.103 & 0.971 & 0.218 \\
S8: & Skills and knowledge & 0.496 & 0.441 & \textcolor{darkred}{\textbf{0.015}} & 0.131 & 0.778 & 0.277 & \textcolor{darkred}{\textbf{0.006}} \\
S9: & Authentic & 0.923 & 0.842 & 0.088 & 0.262 & 0.906 & \textcolor{darkred}{\textbf{0.016}} & 0.141 \\
\hline

S10: & Big leap & 0.342 & 0.385 & 0.939 & 0.489 & 0.676 & 0.808 & 0.276 \\
S11: & Learnt a lot & 0.759 & 0.648 & 0.169 & 0.473 & 0.764 & 0.378 & 0.824 \\
S12: & Authentic & 0.469 & \textcolor{darkred}{\textbf{0.049}} & \textcolor{darkred}{\textbf{0.009}} & 0.347 & 0.698 & \textcolor{darkred}{\textbf{0.026}} & 0.396 \\
S13: & Difference & 0.930 & \textcolor{darkred}{\textbf{0.036}} & 0.327 & 0.791 & 0.167 & 0.533 & 0.759 \\
\hline

S14: & Course structure & 0.860 & 0.549 & 0.122 & 0.452 & 0.805 & 0.213 & 0.497 \\
S15: & Little insight & 0.318 & 0.632 & \textcolor{darkred}{\textbf{0.008}} & 0.312 & 0.916 & 0.089 & 0.597 \\
\hline

S16: & Assessment & 0.628 & \textcolor{darkred}{\textbf{0.023}} & 0.674 & 0.487 & 0.718 & 0.399 & 0.323 \\
S17: & Variation in lab assessment & 0.213 & 0.131 & 0.759 & 0.487 & 0.853 & 0.790 & 0.695 \\
S18: & Variation for project & 0.072 & 0.448 & 0.583 & 0.703 & 0.666 & 0.833 & 0.864 \\
S19: & Variation for help & 0.377 & 0.087 & 0.374 & 0.415 & 0.601 & 0.509 & \textcolor{darkred}{\textbf{0.047}} \\
S20: & Tutorials & 0.539 & 0.053 & 0.691 & 0.206 & 0.425 & 0.950 & 0.400 \\
S21: & Professionalism & 0.531 & 0.667 & \textcolor{darkred}{\textbf{0.006}} & \textcolor{darkred}{\textbf{0.028}} & \textcolor{darkred}{\textbf{0.029}} & 0.194 & \textcolor{darkred}{\textbf{0.013}} \\
S22: & Feedback & \textcolor{darkred}{\textbf{0.018}} & 0.365 & 0.218 & 0.449 & 0.815 & 0.839 & 0.663 \\
S23: & Asking TA help & 0.503 & 0.712 & 0.645 & 0.691 & 0.719 & 0.239 & 0.095 \\
S24: & Preferred internet & \textcolor{darkred}{\textbf{0.024}} & 0.980 & 0.138 & 0.244 & 0.137 & 0.880 & 0.881 \\
S25: & Few TAs & 0.257 & 0.867 & \textcolor{darkred}{\textbf{0.030}} & 0.059 & 0.373 & 0.789 & 0.900 \\
\hline

\end{tabular}
\label{tab:week7_analysis}
\end{adjustbox}
\end{table}

\begin{table}[]
\centering
\caption{Analysis of $p$-values for Week-5}
\begin{adjustbox}{width=0.45\textwidth} 
\begin{tabular}{ll|c|c|c|c|c|c|c}
\hline
\textbf{SN.} & \textbf{Statement} & \textbf{G} & \textbf{PCE} & \textbf{PAE} & \textbf{PCoE} & \textbf{PSE} & \textbf{Gd} & \textbf{B} \\ \hline

S1: & Learning activity & 0.415 & 0.380 & \textcolor{darkred}{\textbf{0.002}} & 0.061 & 0.503 & 0.777 & 0.395 \\ 
S2: & Necessary evil & 0.142 & 0.881 & 0.615 & 0.900 & 0.641 & 0.157 & 0.067 \\ 
S3: & Stressful & 0.866 & 0.608 & 0.488 & 0.898 & 0.610 & 0.504 & 0.709 \\ 
S4: & Help from peers & 0.113 & 0.578 & 0.108 & 0.118 & 0.549 & 0.054 & 0.345 \\
S5: & Help to peers & 0.197 & 0.350 & 0.217 & 0.190 & 0.230 & 0.160 & 0.602 \\ \hline

S6: & Predictable question types & 0.059 & 0.313 & \textcolor{darkred}{\textbf{0.029}} & \textcolor{darkred}{\textbf{0.035}} & \textcolor{darkred}{\textbf{0.011}} & 0.591 & 0.531 \\ 
S7: & Predictable assessment & 0.313 & 0.648 & 0.689 & 0.849 & 0.074 & 0.374 & 0.474 \\ 
S8: & Skills and knowledge & 0.583 & 0.560 & 0.302 & 0.127 & 0.126 & 0.521 & 0.495 \\
S9: & Authentic & 0.621 & 0.321 & 0.125 & 0.745 & 0.626 & 0.533 & 0.087 \\ \hline

S10: & Big leap & 0.785 & 0.150 & 0.406 & 0.543 & 0.941 & 0.902 & 0.104 \\ 
S11: & Learnt a lot & 0.677 & 0.676 & 0.273 & 0.281 & 0.358 & 0.808 & 0.069 \\ 
S12: & Authentic & 0.867 & 0.348 & \textcolor{darkred}{\textbf{0.003}} & \textcolor{darkred}{\textbf{0.020}} & 0.867 & 0.621 & 0.116 \\
S13: & Difference & 0.338 & 0.791 & 0.452 & 0.197 & 0.281 & \textcolor{darkred}{\textbf{0.033}} & \textcolor{darkred}{\textbf{0.021}} \\ \hline

S14: & Course structure & 0.857 & 0.386 & 0.142 & 0.059 & 0.383 & 0.582 & 0.670 \\ 
S15: & Little insight & 0.391 & 0.498 & \textcolor{darkred}{\textbf{0.017}} & 0.092 & 0.201 & \textcolor{darkred}{\textbf{0.005}} & 0.147 \\ \hline

S16: & Assessment & 0.131 & \textcolor{darkred}{\textbf{0.030}} & 0.736 & 0.438 & 0.158 & 0.734 & 0.313 \\ 
S17: & Variation in lab assessment & 0.187 & 0.062 & 0.919 & 0.784 & 0.999 & 0.985 & 0.344 \\ 
S18: & Variation for project & 0.690 & 0.639 & 0.655 & 0.635 & 0.683 & 0.895 & 0.628 \\ 
S19: & Variation for help & 0.463 & 0.872 & 0.940 & 0.785 & 0.283 & 0.843 & 0.263 \\ 
S20: & Tutorials & 0.791 & 0.104 & 0.184 & 0.305 & 0.122 & 0.822 & \textcolor{darkred}{\textbf{0.018}} \\ 
S21: & Professionalism & 0.853 & 0.526 & 0.206 & \textcolor{darkred}{\textbf{0.026}} & \textcolor{darkred}{\textbf{0.037}} & 0.324 & \textcolor{darkred}{\textbf{0.040}} \\ 
S22: & Feedback & 0.905 & 0.845 & 0.438 & 0.288 & 0.933 & 0.836 & 0.162 \\ 
S23: & Asking TA help & 0.932 & 0.920 & 0.360 & 0.346 & 0.240 & 0.366 & 0.276 \\ 
S24: & Preferred internet & 0.487 & 0.698 & 0.327 & 0.155 & 0.105 & 0.641 & 0.813 \\ 
S25: & Few TAs & 0.431 & 0.944 & 0.599 & 0.194 & 0.924 & 0.424 & 0.619 \\ \hline
\end{tabular}
\label{tab:week8_analysis}
\end{adjustbox}
\end{table}

\begin{table}[]
\centering
\caption{Analysis of $p$-values for Week-6}
\begin{adjustbox}{width=0.45\textwidth} 
\begin{tabular}{ll|c|c|c|c|c|c|c}
\hline
\textbf{SN.} & \textbf{Statement} & \textbf{G} & \textbf{PCE} & \textbf{PAE} & \textbf{PCoE} & \textbf{PSE} & \textbf{Gd} & \textbf{B} \\
\hline

S1: & Learning activity & 0.078 & 0.575 & 0.704 & 0.397 & 0.238 & 0.619 & 0.813 \\ 
S2: & Necessary evil & 0.240 & 0.236 & 0.511 & 0.403 & 0.856 & 0.781 & 0.283 \\ 
S3: & Stressful & 0.807 & 0.843 & 0.824 & \textcolor{darkred}{\textbf{0.010}} & 0.662 & 0.959 & 0.611 \\ 
S4: & Help from peers & 0.170 & 0.300 & 0.388 & 0.295 & 0.712 & \textcolor{darkred}{\textbf{0.003}} & 0.301 \\ 
S5: & Help to peers & 0.649 & 0.948 & 0.678 & 0.750 & 0.760 & 0.586 & 0.296 \\ \hline

S6: & Predictable question types & \textcolor{darkred}{\textbf{0.020}} & 0.299 & 0.085 & 0.938 & 0.689 & 0.179 & 0.090 \\ 
S7: & Predictable assessment & 0.075 & 0.307 & 0.082 & 0.847 & 0.219 & 0.614 & 0.531 \\
S8: & Skills and knowledge & 0.517 & 0.060 & 0.396 & 0.452 & \textcolor{darkred}{\textbf{0.019}} & 0.353 & 0.526 \\ 
S9: & Authentic & 0.405 & 0.309 & 0.125 & 0.765 & \textcolor{darkred}{\textbf{0.009}} & 0.415 & 0.901 \\ \hline

S10: & Big leap & \textcolor{darkred}{\textbf{0.018}} & 0.803 & 0.427 & 0.851 & 0.484 & 0.517 & 0.630 \\ 
S11: & Learnt a lot & 0.784 & 0.565 & 0.866 & 0.314 & 0.104 & 0.520 & 0.695 \\ 
S12: & Authentic & 0.858 & 0.063 & 0.158 & 0.212 & 0.393 & 0.573 & 0.672 \\ 
S13: & Difference & 0.295 & 0.588 & 0.929 & 0.127 & 0.140 & 0.370 & 0.604 \\ \hline

S14: & Course structure & 0.175 & 0.239 & 0.501 & 0.366 & 0.562 & 0.801 & 0.636 \\
S15: & Little insight & 0.075 & 0.683 & 0.203 & 0.377 & 0.270 & 0.239 & 0.797 \\ \hline

S16: & Assessment & 0.471 & 0.179 & 0.272 & 0.344 & 0.795 & 0.180 & 0.592 \\ 
S17: & Variation in lab assessment & 0.152 & 0.778 & \textcolor{darkred}{\textbf{0.033}} & 0.221 & \textcolor{darkred}{\textbf{0.046}} & 0.760 & 0.263 \\ 
S18: & Variation for project & \textcolor{darkred}{\textbf{0.012}} & 0.130 & 0.726 & 0.427 & 0.134 & 0.925 & 0.696 \\ 
S19: & Variation for help & 0.410 & 0.330 & 0.118 & 0.546 & 0.569 & 0.860 & 0.622 \\ 
S20: & Tutorials & \textcolor{darkred}{\textbf{0.038}} & 0.622 & 0.437 & 0.062 & 0.058 & 0.812 & 0.556 \\ 
S21: & Professionalism & 0.251 & 0.492 & 0.356 & 0.880 & 0.812 & 0.962 & 0.100 \\ 
S22: & Feedback & 0.304 & 0.265 & \textcolor{darkred}{\textbf{0.020}} & 0.258 & \textcolor{darkred}{\textbf{0.046}} & 0.887 & 0.899 \\ 
S23: & Asking TA help & 0.057 & \textcolor{darkred}{\textbf{0.002}} & 0.916 & 0.764 & \textcolor{darkred}{\textbf{0.045}} & 0.715 & 0.684 \\ 
S24: & Preferred internet & 0.639 & 0.578 & \textcolor{darkred}{\textbf{0.005}} & 0.118 & 0.063 & 0.884 & 0.226 \\
S25: & Few TAs & 0.099 & 0.120 & 0.184 & 0.753 & 0.686 & 0.962 & 0.912 \\ \hline

\end{tabular}
\label{tab:week9_analysis}
\end{adjustbox}
\end{table}

\begin{table}[]
\centering
\caption{Analysis of $p$-values for Week-7}
\begin{adjustbox}{width=0.45\textwidth} 
\begin{tabular}{ll|c|c|c|c|c|c|c}
\hline
\textbf{SN.} & \textbf{Statement} & \textbf{G} & \textbf{PCE} & \textbf{PAE} & \textbf{PCoE} & \textbf{PSE} & \textbf{Gd} & \textbf{B} \\ \hline

S1: & Learning activity & 0.788 & 0.231 & 0.568 & \textcolor{darkred}{\textbf{0.003}} & 0.086 & 0.149 & 0.984 \\ 
S2: & Necessary evil & 0.714 & \textcolor{darkred}{\textbf{0.039}} & 0.537 & 0.207 & 0.581 & 0.199 & 0.246 \\ 
S3: & Stressful & 0.438 & 0.561 & 0.253 & 0.944 & 0.712 & 0.399 & 0.878 \\ 
S4: & Help from peers & 0.347 & 0.125 & 0.358 & 0.118 & 0.707 & 0.118 & 0.183 \\
S5: & Help to peers & 0.799 & 0.777 & 0.084 & 0.436 & 0.893 & 0.077 & 0.348 \\ \hline

S6: & Predictable question types & 0.562 & 0.691 & 0.663 & 0.575 & 0.126 & 0.060 & 0.813 \\ 
S7: & Predictable assessment & 0.687 & 0.780 & 0.323 & 0.834 & 0.132 & 0.826 & 0.679 \\ 
S8: & Skills and knowledge & 0.471 & 0.892 & \textcolor{darkred}{\textbf{0.047}} & 0.728 & 0.484 & 0.596 & 0.592 \\
S9: & Authentic & 0.313 & 0.574 & 0.097 & 0.506 & 0.397 & 0.083 & 0.543 \\ \hline

S10: & Big leap & 0.942 & 0.154 & 0.971 & 0.506 & 0.714 & 0.638 & 0.291 \\ 
S11: & Learnt a lot & 0.231 & \textcolor{darkred}{\textbf{0.023}} & 0.214 & 0.675 & \textcolor{darkred}{\textbf{0.014}} & \textcolor{darkred}{\textbf{0.027}} & 0.435 \\ 
S12: & Authentic & 0.544 & \textcolor{darkred}{\textbf{0.041}} & 0.281 & 0.096 & 0.980 & 0.273 & 0.681 \\ 
S13: & Difference & 0.897 & 0.391 & 0.605 & 0.781 & 0.334 & 0.837 & 0.098 \\ \hline

S14: & Course structure & 0.874 & 0.170 & 0.456 & \textcolor{darkred}{\textbf{0.003}} & 0.196 & 0.531 & 0.413 \\
S15: & Little insight & 0.752 & 0.319 & \textcolor{darkred}{\textbf{0.014}} & \textcolor{darkred}{\textbf{0.008}} & \textcolor{darkred}{\textbf{0.041}} & \textcolor{darkred}{\textbf{0.008}} & 0.512 \\ \hline

S16: & Assessment & 0.485 & 0.167 & 0.799 & 0.657 & 0.487 & 0.623 & 0.596 \\ 
S17: & Variation in lab assessment & 0.304 & 0.190 & 0.124 & 0.384 & 0.530 & 0.701 & 0.790 \\ 
S18: & Variation for project & 0.941 & 0.115 & 0.925 & 0.840 & 0.922 & 0.665 & 0.625 \\ 
S19: & Variation for help & 0.851 & 0.228 & 0.961 & 0.256 & 0.205 & 0.322 & 0.095 \\ 
S20: & Tutorials & 0.325 & 0.164 & 0.930 & 0.788 & 0.926 & 0.075 & 0.058 \\ 
S21: & Professionalism & 0.949 & 0.978 & 0.361 & 0.197 & 0.898 & 0.614 & 0.812 \\ 
S22: & Feedback & 0.351 & 0.215 & 0.152 & 0.193 & 0.070 & 0.241 & 0.950 \\

S23: & Asking TA help & 0.432 & \textcolor{darkred}{\textbf{0.039}} & 0.746 & 0.875 & 0.298 & 0.747 & \textcolor{darkred}{\textbf{0.044}} \\ 
S24: & Preferred internet & 0.686 & 0.070 & 0.358 & 0.175 & 0.263 & 0.780 & 0.974 \\ 
S25: & Few TAs & 0.663 & 0.078 & 0.252 & 0.136 & 0.367 & 0.443 & 0.175 \\ \hline
\end{tabular}
\label{tab:week10_analysis}
\end{adjustbox}
\end{table}



\begin{table}[]
\centering
\caption{Analysis of $p$-values for Week-8}
\begin{adjustbox}{width=0.45\textwidth} 
\begin{tabular}{ll|c|c|c|c|c|c|c}
\hline
\textbf{SN.} & \textbf{Statement} & \textbf{G} & \textbf{PCE} & \textbf{PAE} & \textbf{PCoE} & \textbf{PSE} & \textbf{Gd} & \textbf{B} \\ \hline

S1: & Learning activity       & 0.927 & 0.947 & 0.230 & 0.118 & 0.087 & 0.081 & 0.938 \\
S2: & Necessary evil          & 0.635 & 0.305 & 0.820 & 0.347 & 0.166 & 0.534 & 0.120 \\ 
S3: & Stressful               & 0.054 & 0.105 & 0.746 & 0.078 & 0.554 & 0.547 & 0.655 \\ 
S4: & Help from peers    & 0.254 & \textcolor{darkred}{\textbf{0.040}} & 0.595 & 0.678 & 0.366 & \textcolor{darkred}{\textbf{0.022}} & 0.877 \\
S5: & Help to peers            & 0.572 & 0.958 & 0.630 & 0.144 & 0.294 & 0.519 & 0.983 \\ \hline

S6: & Predictable question types & 0.640 & 0.655 & 0.568 & 0.654 & 0.673 & \textcolor{darkred}{\textbf{0.045}} & 0.251 \\ 
S7: & Predictable assessment  & 0.299 & 0.564 & 0.836 & 0.783 & 0.938 & 0.113 & 0.563 \\ 
S8: & Skills and knowledge & 0.244 & 0.444 & 0.739 & 0.385 & 0.407 & 0.456 & 0.712 \\ 
S9: & Authentic               & 0.578 & 0.677 & 0.463 & 0.172 & 0.237 & 0.224 & \textcolor{darkred}{\textbf{0.027}} \\ \hline

S10: & Big leap     & 0.109 & 0.259 & 0.332 & 0.415 & \textcolor{darkred}{\textbf{0.037}} & 0.362 & 0.737 \\ 
S11: & Learnt a lot           & 0.661 & 0.610 & 0.342 & 0.593 & \textcolor{darkred}{\textbf{0.034}} & 0.409 & 0.375 \\ 
S12: & Authentic     & 0.880 & 0.156 & 0.590 & 0.113 & 0.341 & 0.691 & 0.504 \\ 
S13: & Difference & 0.071 & 0.508 & 0.615 & 0.223 & 0.574 & 0.178 & 0.850 \\ \hline

S14: & Course structure    & 0.955 & 0.285 & 0.101 & 0.177 & 0.100 & 0.818 & 0.525 \\ 
S15: & Little insight      & 0.236 & 0.973 & \textcolor{darkred}{\textbf{0.023}} & 0.157 & 0.654 & \textcolor{darkred}{\textbf{0.014}} & \textcolor{darkred}{\textbf{0.048}} \\ \hline
S16: & Assessment & 0.760 & 0.325 & 0.632 & 0.199 & 0.510 & 0.681 & 0.980 \\ 
S17: & Variation in lab assessment & 0.893 & 0.424 & 0.671 & 0.968 & 0.571 & 0.863 & 0.567 \\ 
S18: & Variation for project & 0.617 & 0.129 & 0.267 & 0.220 & 0.751 & 0.255 & 0.527 \\ 
S19: & Variation for help         & 0.847 & 0.313 & 0.449 & 0.450 & 0.423 & 0.162 & 0.362 \\ 
S20: & Tutorials  & 0.332 & 0.561 & 0.420 & 0.628 & 0.474 & 0.055 & \textcolor{darkred}{\textbf{0.043}} \\ 
S21: & Professionalism        & 0.950 & 0.842 & 0.933 & 0.852 & 0.560 & 0.273 & 0.245 \\ 
S22: & Feedback        & 0.391 & 0.958 & 0.420 & 0.940 & 0.129 & 0.573 & 0.598 \\ 
S23: & Asking TA help         & 0.195 & 0.063 & 0.769 & 0.634 & 0.531 & 0.700 & 0.507 \\ 
S24: & Preferred internet & 0.120 & 0.374 & 0.062 & 0.257 & 0.093 & 0.811 & 0.767 \\ 
S25: & Few TAs    & 0.952 & \textcolor{darkred}{\textbf{0.035}} & 0.379 & 0.851 & 0.114 & 0.429 & 0.308 \\ \hline
\end{tabular}
\label{tab:week11_analysis}
\end{adjustbox}
\end{table}



\end{document}